\documentclass[apj,iop,twocolappendix]{emulateapj}
\usepackage{natbib}
\usepackage{float}
\usepackage{graphicx}
\usepackage{epstopdf}
\usepackage{amsmath}
\usepackage{mathrsfs}
\usepackage{textcomp}

\begin{document}

\shorttitle{The Thermal Proximity Effect}
\shortauthors{Khrykin et al.}

\title{The Thermal Proximity Effect: A new Probe of the \ion{He}{2} Reionization History and the Quasar Lifetime}

\author{I.S.~Khrykin\altaffilmark{1,2,3}\footnotemark[*], J.F.~Hennawi\altaffilmark{1,4}, M.~McQuinn\altaffilmark{5}}

\footnotetext[*]{e-mail: khrykin@mpia-hd.mpg.de}
\altaffiltext{1}{ Max-Planck-Institut f{\"u}r Astronomie, K\"onigstuhl 17, D-69117 Heidelberg, Germany}
\altaffiltext{2}{International Max Planck Research School for Astronomy \& Cosmic Physics at the University of Heidelberg, K\"onigstuhl 17, D-69117 Heidelberg, Germany}
\altaffiltext{3}{ Southern Federal University, Stachki Avenue 194, 344090 Rostov-on-Don, Russian Federation}
\altaffiltext{4}{ Department of Physics, University of California, Santa Barbara, CA 93106, USA}
\altaffiltext{5}{ University of Washington, Department of Astronomy, 3910 15th Ave NE, WA 98195-1580 Seattle, USA}

\begin{abstract}

Despite decades of effort, the timing and duration of \ion{He}{2} reionization and the properties of the quasars believed to drive it, are still not well constrained. We present a new method to study both via the \emph{thermal proximity effect} -- the heating of the intergalactic medium (IGM) around quasars when their radiation doubly ionizes helium. We post-process hydrodynamical simulations with $1$D radiative transfer and study how the thermal proximity effect depends on \ion{He}{2} fraction, $x_{\rm HeII,0}$, which prevailed in the IGM before the quasar turned on, and the quasar lifetime $t_{\rm Q}$.
We find that the amplitude of the temperature boost in the quasar environment depends on $x_{\rm HeII,0}$, with a characteristic value of $\Delta T\simeq 10^4\,{\rm K}$ for $x_{\rm HeII,0} = 1.0$, whereas the size of the thermal proximity zone is sensitive to $t_{\rm Q}$,with typical sizes of $\simeq100\,{\rm cMpc}$ for $t_{\rm Q}=10^8\,{\rm yr}$. This temperature boost increases the thermal broadening of \ion{H}{1} absorption lines near the quasar. We introduce a new Bayesian statistical method based on measuring the Ly$\alpha$ forest power spectrum as a function of distance from the quasar, and demonstrate that the thermal proximity effect should be easily detectable. For a mock dataset of $50$ quasars at $z\simeq4$, we predict that one can measure $x_{\rm HeII,0}$ to an (absolute) precision $\approx0.04$, and $t_{\rm Q}$ to a precision of $\approx0.1$~dex. By applying our formalism to existing high-resolution Ly$\alpha$ forest spectra, one should be able to reconstruct the \ion{He}{2} reionization history,providing a global census of hard photons in the high-$z$ universe.

\end{abstract}

\begin{keywords}
{cosmology: theory -- dark ages, reionization, first stars -- intergalactic medium -- quasars: general}
\end{keywords}

\maketitle

\section{Introduction}

The Epoch of Reionization, when the intergalactic medium was ionized by astrophysical sources, is a key juncture in the history of the Universe.  When and how reionization occurred informs models for the formation and evolution of the first stars, galaxies, and large-scale structure.
Constraints on the Epoch of Reionization are derived mainly from the measurements of the Thompson scattering optical depth from the Cosmic Microwave Background \citep{Robertson2015, PC2016} together with studies of the Ly$\alpha$ forest opacity towards high-redshift quasars \citep{Fan2006, McGreer2011}. It is widely believed that galaxies provide enough ionizing photons \citep{Robertson2010, Finkelstein2012} to reionize intergalactic hydrogen by $z \simeq 6$. However, radiation from galaxies is unlikely to have been hard enough to have doubly ionized helium (\ion{He}{2} $\longrightarrow$ \ion{He}{3}, requiring $h\nu \ge 54.4$~eV). The current paradigm is that the complete reionization of helium was delayed until lower redshifts ($z \sim 3$), when quasars, which can emit the required energetic photons, became sufficiently abundant  \citep{Madau1994, Miralda2000, McQuinn2009, Haardt2012, Compostella2013, Compostella2014, LaPlante2016}. 

The focus of this paper is on a new method to constrain this reionization of the last electron of helium, termed  \ion{He}{2} reionization.  The temporal extent and morphology of \ion{He}{2} reionization remains relatively unconstrained. Intergalactic \ion{He}{2} Ly$\alpha$ absorption ($\lambda_{{\rm Ly}\alpha}^{\rm rest} = 303.78 {\rm \AA}$) in the far-UV spectra of $z_{\rm em} \simeq 3-4$ quasars directly probes \ion{He}{2} in the IGM  \citep{Hogan1997, Anderson1999, Heap2000, Shull2010, Worseck2011, Syphers2012, Syphers2014, Worseck2014, Zheng2015}. The recent discovery of regions in the IGM at $z \gtrsim 3.3$ with significant
transmission \citep{Worseck2011, Worseck2014} suggests that \ion{He}{2} reionization may have occurred earlier than current models predict
(which generically reionize  \ion{He}{2} at the peak of the quasar epoch at $z \simeq 3$; \citealt{McQuinn2009, Compostella2013, Compostella2014, LaPlante2016}).  However, this discrepancy could also result from uncertainties in the simulations and in simplistic models for quasar light curves \citep{DAloisio2016}.

If \ion{He}{2} reionization did occur significantly earlier than
current models predict, additional photons beyond those produced by
$\gtrsim L_*$ quasars might be required. A possible candidate is a
population of faint quasars \citep{Giallongo2015, Madau2015}, but
other exotic sources have also been proposed, including UV emission
from $\sim10^6$\,K halos \citep{Miniati2004}, primordial globular
clusters \citep{Power2009}, mini-quasars \citep{Madau2004}, and dark
matter annihilations \citep{Araya2014}. Irrespective of the nature of
these sources, if they reionized intergalactic helium at $z \gtrsim
4$, these sources also likely contributed substantially to \ion{H}{1}
reionization \citep{Madau2015}. Unfortunately, it is not currently
possible to probe the ionization state of \ion{He}{2} via
far-UV \ion{He}{2} Ly$\alpha$ absorption at $z \gtrsim 4$, and constraints on
\ion{He}{2} reionization from the thermal state of the IGM do not all agree
\citep{Lidz2010,Becker2011,Upton2016}. This paper
introduces a new method to constrain the timing of \ion{He}{2} reionization.

In the last decades significant effort was invested in measuring the thermal state of the IGM, aiming to constrain the timing of \ion{H}{1} and \ion{He}{2} reionization epochs. During these reionization events ionization fronts propagated supersonically impulsively heating IGM gas to $\sim 10^4~{\rm K}$, though the exact amount of injected heat depends on the spectral shape and abundance of the ionizing sources, and the opacity of the IGM \citep{McQuinn2012, Davies2016}. Hydrodynamical simulations show that within several hundred Myr of reionization, IGM gas relaxes onto a tight power law temperature-density relation $T = T_0(\rho\slash {\bar \rho})^{\gamma-1}$, where $T_0$ is the temperature at the cosmic mean density, $\bar \rho$ \citep{Hui1997, McQuinn2016}. However, quasar driven \ion{He}{2} reionization, which is thought to increase the temperature of the IGM by $\Delta T \simeq 5-10 \times
10^{3}$~K at $z \simeq 3-4$ \citep{McQuinn2009, Compostella2013}, can
lead to a significant changes in this relation \citep{McQuinn2009,
  Upton2016, LaPlante2016}. A comparable temperature increase at $z \simeq 3$ was reported
by \citet{Schaye2000} (see also \citealp{Theuns2002c}), which was
later qualitatively supported by \citet{Becker2011}. However, there is
generally a lack of consistency between these measurements and those from other
studies \citep{Ricotti2000, McDonald2001, Zaldarriaga2001,
  Lidz2010}. In particular \citet{Lidz2010} found evidence for a much
hotter IGM at $z \gtrsim 3.5$ than expected from current models
\citep{Puchwein2015, Upton2016, Onorbe2016}, in disagreement with
\citet{Becker2011}.

Another important ingredient in reionization models is the quasar
lifetime, which determines the morphology of the hard UV background
\citep{McQuinn2009, Compostella2013, Compostella2014,McQuinn2014}.
Recent lifetime estimates based on the light travel time arguments in quasar host galaxies \citep[$t_{\rm Q}\simeq 10^5$~yr;][]{Schawinski2010, Schawinski2015}, and quasar powered Ly$\alpha$ fluorescence \citep[$10^6 \lesssim t_{\rm Q} \lesssim 3\times 10^7$~yr;][]{Trainor2013, Borisova2015} are indirect, and alternative physical mechanisms can be invoked to explain the observations. To date the most robust estimates of the quasar lifetime are derived from the line-of-sight \ion{H}{1} proximity effect \citep{Bajtlik1988, HC2002} in Ly$\alpha$ forest spectra. The quasar must shine for $t_{\rm Q} \gtrsim t_{\rm
  eq}$ to produce a detectable proximity zone, where the equilibration timescale $t_{\rm eq} \simeq
\Gamma_{\rm HI}^{-1}$ is the time it takes for the IGM to reach ionization
equilibrium with the quasar radiation. At $z \simeq 3-5$, the \ion{H}{1} photoionization rate is constrained to be $\Gamma_{\rm HI} \simeq 10^{-12}{\rm s^{-1}}$, which means light curve variations from the proximate quasar will only produce a line-of-sight proximity effect if $t_{\rm Q} \gtrsim 3 \times 10^4$~yr.  On a positive note, \citet{Khrykin2016} demonstrated that \ion{He}{2} Ly$\alpha$ line-of-sight proximity zones in the far-UV spectra of $z \simeq 3-4$ can probe quasar lifetimes on the more interesting timescale of $\sim 10$~Myr, although this method has yet to be applied to real data. Here we propose another method that we show is sensitive to comparable lifetimes based on \ion{H}{1} Ly$\alpha$ spectra, for which there exists a substantial amount of high-${\rm S\slash N}$, high-resolution data. 

Before reionization is complete, quasars that turn on in a medium in which the helium is initially \ion{He}{2}, doubly ionize it, creating \ion{He}{3} regions. Photoionization of \ion{He}{2} $\longrightarrow$ \ion{He}{3} produces suprathermal photoelectrons, which are thermalized, raising the temperature of the gas by $\sim 10^4$K \citep{Abel1999, Miralda1994, McQuinn2009, Bolton2009}. Such an increase in IGM temperatures in the quasar environs is referred to as the \emph{thermal proximity effect}. As discussed by \citet{Bolton2009} and \citet{Meiksin2010}, this temperature boost due to the evolution of \ion{He}{2} fraction can significantly alter the \ion{H}{1} Ly$\alpha$ forest around the quasar via additional thermal broadening of the Ly$\alpha$ absorption lines. For instance, \citet{Bolton2010,Bolton2012} claimed to measure IGM temperatures $\simeq 0.3$~dex higher than expected in the proximity zones of $z \sim 6$ quasars, attributing this extra heating to the quasar reionizing its surrounding \ion{He}{2}. They also showed that this temperature boost can be used to infer the quasar lifetime, and estimated a lower limit $t_{\rm Q} \gtrsim 10^{6.5}$~yr. 

The aim of this paper is to further investigate the line-of-sight thermal proximity effect, quantify its detectability, and determine how well
it can constrain the timing of \ion{He}{2} reionization and the
average quasar lifetime.  We combine cosmological hydrodynamical
simulations with post-processing radiative transfer calculations using
the $1$D radiative transfer algorithm specifically developed in
\citet{Khrykin2016} for this purpose, which allows us to track the
changes in the ionization state of helium and hydrogen, and the
associated temperature of the IGM around quasars. We apply the power
spectrum statistic to the output of our simulations, which is directly
sensitive to thermal broadening, and thus to the average temperature
profile around quasars. We use Bayesian methods with Markov Chain
Monte Carlo (MCMC) calculations to determine the accuracy with which
the amount of singly ionized helium in the IGM and the quasar lifetime
can be determined.

This paper is organized as follows. In \S~\ref{sec:model} we briefly
explore the most important parameters of our numerical model. We
describe the thermal proximity effect during \ion{He}{2} reionization
and show its impact on the properties of the IGM gas in
\S~\ref{sec:T_IGM}. We introduce our method for measuring
\ion{He}{2} fraction  and quasar 
lifetime from the $1$D Ly$\alpha$ forest
power spectrum in \S~\ref{sec:powspec}. We present the
results of our MCMC analysis in \S~\ref{sec:est}, and explore
the possibility of reconstructing the history \ion{He}{2} reionization in
\S~\ref{sec:history}. We briefly explore wavelet analysis as another method to
detect the thermal proximity effect in \S~\ref{sec:wavelets}. Possible uncertainties and systematic errors are discussed in \S~\ref{sec:discussion}, and we
summarize and conclude in \S~\ref{sec:conclusions}.

Throughout this paper we assume a flat $\Lambda$CDM cosmology with Hubble constant $h=0.7$, $\Omega_{\rm m}=0.27$, $\Omega_{\rm b}=0.046$, $\sigma_8=0.8$ and $n_s = 0.96$, and helium mass fraction $Y_{\rm He}=0.24$, consistent with recent Planck results \citep{PC2016}. All distances are quoted in units of comoving Mpc, i.e., cMpc. 

As noted above, one of the goals of this paper is to understand the
constraints that can be put on the quasar \emph{lifetime}, which is
the time spanned by a single episode of accretion onto the black
hole. One should distinguish this from the quasar \emph{duty cycle},
which refers to the total time that galaxies shine as active
quasars. In the context of proximity effects in the IGM, one actually
only constrains the quasar \emph{on-time}, which we will denote as
$t_{\rm Q}$. If we imagine that time $t=0$ corresponds to the time
when the quasar emitted light that is just now reaching our telescopes
on Earth, then the quasar on-time is defined such that the quasar
turned on at time $-t_{\rm Q}$ in the past.  This timescale is, in
fact, a lower limit on the quasar lifetime, which arises from the fact
that we observe a proximity zone at $t=0$ when the quasar has been
shining for time $t_{\rm Q}$, whereas this quasar episode may indeed
continue, which we can only record on Earth if we could conduct
observations in the future.  For simplicity in the text, we will
henceforth refer to the quasar on-time as the \emph{quasar lifetime}
denoted by $t_{\rm Q}$, but the reader should always bear in mind that
this is actually a lower limit on the duration of quasar emission episodes.

\section{Summary of the Numerical model}
\label{sec:model}

We use a combination of smooth particle hydrodynamics (SPH)
simulations and a $1$D post-processing radiative transfer algorithm to
study the thermal evolution of the intergalactic medium around quasars. 
In this section we present the most important features of our model 
and refer the reader to the more detailed
description given in \citet{Khrykin2016}.

\subsection{Hydrodynamical Simulations}

We run SPH simulations using the Tree-SPH code Gadget-3 \citep{Springel2005} with $2 \times 512^3$ particles and a box size of $25h^{-1}$~cMpc. Starting at the location of the most massive halos (${\rm M} > 5 \times 10^{11} {\rm M}_{\odot}$) at $z = \left[3.1, 3.5, 3.9, 4.3, 4.6 \right]$ we extract $1000$ sightlines (which we will refer to  as \emph{skewers}) by casting the rays through the simulation box at random angles and
using the periodic boundary conditions to wrap the skewers through the
simulation volume. The resulting skewers have a total length of
$160$~cMpc with a pixel scale of ${\rm d}r = 0.01$~cMpc (${\rm d}v =
1.0\ {\rm km\ s^{-1}}$). While most of the discussion in this paper
focuses on the output at $z = 3.9$, we consider other redshifts as
well in \S~\ref{sec:history}. Note that we also perform our
calculations at redshift $z = 5.0$, for which we do not have outputs
of the hydrodynamical simulations. However, assuming that there is no evolution of the density field between two redshifts, and the only change is due to the cosmological expansion, we simply re-scaled the
density field of $z = 4.6$ output by a factor of $\left( 1 +
z\right)^{3}$, which is a good approximation.

\subsection{Radiative Transfer Algorithm}
\label{sec:rt_lum}

Skewers are extracted from the SPH simulation volume, and are post-processed using our $1$D radiative transfer algorithm, based on the C$^2$-Ray
algorithm \citep{Mellema2006}. We assume the quasar, placed at the beginning of each skewer, has a spectral energy distribution (SED) which can be
approximated by a power-law, such that its specific photon production rate $N_{\nu}$ at frequencies $\nu \geq \nu_{\rm th}$ is given by
\begin{equation}\label{eqn:lum}
N_{\nu} = \frac{\alpha Q}{\rm \nu_{th}} \left(\frac{\rm \nu}{\nu_{\rm th}}\right)^{-\left(\alpha + 1\right)}, 
\end{equation} where $Q$ is the production rate (${\rm photons}\ {\rm s}^{-1}$) of photons
with frequency above the \ion{H}{1} (or \ion{He}{2}) ionization threshold $\nu_{\rm th}$, and  the spectral index is assumed to be $\alpha = 1.5$ ($f_{\nu} \sim \nu^{-\alpha}$), consistent with inferred values from UV composite quasar spectra \citep{Telfer2002,Shull2012,Lusso2015}. Following the procedure described in \citet{Hennawi2006} (see also \citet{Khrykin2016} for details), we find that for a median $i$-band magnitude $i = 18.6$ of quasars at $z \simeq 4$ the $Q_{\rm 1Ry} \simeq 10^{57.4} {\rm s^{-1}}$ and $Q_{\rm 4Ry} \simeq 10^{56.5}{\rm s^{-1}}$, which we adopt as our fiducial values throughout the paper.

\subsubsection{Photoionization of \ion{He}{2}}

The quasar \ion{He}{2} photoionization rate in each cell is given by
\begin{equation}\label{eqn:gamma_qso}
\Gamma_{\rm HeII}^{\rm QSO} = \int_{\nu_{\rm 4Ry}}^{\infty}\frac{N_{\rm
    \nu}e^{-\langle\tau_{\rm
      \nu}\rangle}\left(1-e^{-\langle\delta\tau_{\rm
      \nu}\rangle}\right)}{\langle n_{\rm HeII}\rangle V_{\rm
    cell}}d\nu,
\end{equation}
where $\langle\tau_{\nu}\rangle$ is the optical depth along the skewer from the source to the current cell, $\langle\delta\tau_{\nu}\rangle$ is the optical depth inside the cell, $\langle n_{\rm HeII} \rangle$ is the average number density of \ion{He}{2} in this cell, and $V_{\rm cell}$ is the volume of the cell. The angular brackets indicate time averages over the discrete time step $\delta t$. Combining with eqn.~(\ref{eqn:lum}), eqn.~(\ref{eqn:gamma_qso}) becomes:
\begin{equation}\label{eqn:gammaqso}
\small
\Gamma_{\rm HeII}^{\rm QSO} = \frac{\alpha Q_{\rm 4Ry}}{n_{\rm HeII}V_{\rm cell}\nu_{\rm th}}\int_{\nu_{\rm th}}^{\infty}\left(\frac{\nu}{\nu_{\rm th}}\right)^{-(\alpha + 1)} e^{-\langle\tau_{\rm \nu}\rangle}\left(1-e^{-\langle\delta\tau_{\nu}\rangle}\right)d\nu.
\end{equation}
We note that our radiative transfer algorithm is not tracking multiple frequencies, but rather calculates the frequency-integrated quasar photoionization rate given by eqn~(\ref{eqn:gammaqso}). For a given quasar SED (eqn.~\ref{eqn:lum}), and approximating the \ion{He}{2} absorption cross-section as $\sigma_{\nu} \approx \sigma_{\rm th}\left(\nu \slash \nu_{\rm th} \right)^{-3}$,
eqn.~(\ref{eqn:gammaqso}) has a direct analytic solution depending on $\langle \tau_{\rm th}\rangle$ and $\langle \delta\tau_{\rm th}\rangle$, evaluated at a single frequency corresponding to the \ion{He}{2} $4$~Ry ionization threshold.

In order to account for additional ionizations caused by \ion{He}{2} ionizing photons emitted by other sources in the Universe, we also include the \ion{He}{2} intergalactic ionizing background in our calculations. We approximate it as being constant in space and time, and we add it in each pixel of our skewers, yielding the total \ion{He}{2} photoionization rate $\Gamma_{\rm HeII}^{\rm tot} = \Gamma_{\rm HeII}^{\rm QSO} + \Gamma_{\rm HeII}^{\rm bkg}$. We also include a prescription for more careful modeling of the attenuation of \ion{He}{2} ionizing background by \ion{He}{2}-LLSs \citep{McQuinn2010}, details of which can be found in Appendix B of \citet{Khrykin2016}.  

\subsubsection{Time-evolution of \ion{He}{2} Fraction}
\label{sec:He2f}

Given eqn.~(\ref{eqn:gammaqso}), the time-evolution of the \ion{He}{2} fraction $x_{\rm HeII}\left( t\right)$ is described by
\begin{equation}\label{eqn:dxHeIIdt}
  \frac{dx_{\rm HeII}}{dt} = -\Gamma_{\rm HeII}^{\rm tot}\left(t\right) \, x_{\rm HeII} + n_{\rm e} \alpha_{\rm A}^{\left[ {\rm HeII}\right]} \left (1- x_{\rm HeII} \right),
\end{equation}
where $n_{\rm e}$ is the number density of free electrons, and $\alpha_{\rm A}^{\left[ {\rm HeII}\right]}$ is the \ion{He}{2} Case A recombination coefficient. In \citet{Khrykin2016} we showed that evolution of \ion{He}{2} proximity zones around quasars can be described by the \ion{He}{2} fraction $x_{\rm HeII}$ approaching ionization equilibrium. This approach depends on the characteristic timescale on which the \ion{He}{2} ionization state of IGM gas responds to the changes in the radiation field, i.e., the equilibration time, given by
\begin{equation}\label{eqn:eqtime}
  t_{\rm eq} = \left( \Gamma_{\rm HeII}^{\rm tot} + n_{\rm e}\alpha_{\rm A}^{\left[ {\rm HeII}\right]}\right)^{-1} \approx \Gamma_{\rm HeII}^{\rm tot}\ ^{-1} 
\end{equation}
Since for \ion{He}{2} at $z \simeq 3-4$ the equilibration timescale is $t_{\rm eq}^{\rm \left[ HeII \right]} \approx 10^7$~yr,
comparable to quasar lifetimes $t_{\rm Q}$, the ionization equilibrium is a poor approximation. Hence, we follow the full time-dependent radiative transfer of \ion{He}{2} in the proximity zone.

The mean initial \ion{He}{2} fraction $x_{\rm HeII,0}$ is calculated from $100$ skewers by running our radiative transfer algorithm with quasar turned off, and the \ion{He}{2} background being the only source of ionizations. The desired value of $x_{\rm HeII,0}$ is then set by adjusting the \ion{He}{2} ionizing background. These \ion{He}{2} background values are then used in radiative transfer calculations with quasar turned on.

Further, using our $1$D radiative transfer algorithm we integrate eqn.~(\ref{eqn:dxHeIIdt}) for the time-evolution of $x_{\rm HeII}$ over time $t = t_{\rm Q}$, which denotes the quasar lifetime, yielding the \ion{He}{2} fraction $x_{\rm HeII}$ at each location along the given skewer. We refer the interested reader to \citet{Khrykin2016} for more detailed description of our radiative transfer calculations, the time-evolution of \ion{He}{2} fraction, and its effect on the structure of the quasar proximity zones.

\subsubsection{Treatment of \ion{H}{1}}

Analogous to eqn.~(\ref{eqn:gamma_qso}), our radiative transfer algorithm calculates the quasar \ion{H}{1} photoionization rate $\Gamma_{\rm HI}^{\rm QSO}$ given by
\begin{equation}
\Gamma_{\rm QSO}^{\rm HI} = \frac{1}{4\pi r^2}\int_{\nu_{\rm 1Ry}}^{\infty} N_{\nu}\sigma_{\rm HI}{\rm d}\nu,
\end{equation}
where $\sigma_{\rm HI}$ is \ion{H}{1} absorption cross-section, and we assumed that hydrogen is highly ionized
and hence optically thin to ionizing radiation, which is valid at $z \lesssim 6$, well after \ion{H}{1} reionization has been completed. We also assume that all radiation at frequencies $\nu \gtrsim \nu_{4 {\rm Ry}}$ is absorbed by \ion{He}{2}, which is valid for the same case of highly ionized \ion{H}{1}.

Similar to helium, we introduce the metagalactic \ion{H}{1} ionizing
background, described by the photoionization rate $\Gamma_{\rm
  HI}^{\rm bkg}$, to quantify the influence of all other sources of
\ion{H}{1} ionizing photons, i.e., $\Gamma_{\rm HI}^{\rm tot} =
\Gamma_{\rm HI}^{\rm QSO} + \Gamma_{\rm HI}^{\rm bkg}$. The value of
$\Gamma_{\rm HI}^{\rm bkg}$ is chosen to match the mean transmission at all redshifts considered in this work, consistent with the measurements of
\citet{Becker2013a}\footnote{We match these values of the mean transmitted flux by running radiative transfer simulations with quasar turned off and \ion{H}{1} ionizing background $\Gamma_{\rm HI}^{\rm bkg}$ as the only source.}

Because the \ion{H}{1} equlibration time is so short, i.e. $t_{\rm eq}^{\rm \left[ HI\right]}\sim \Gamma_{\rm HI}^{-1} \approx 10^4$~yr, we assume that \ion{H}{1} is in ionization equilibrium, which is an excellent approximation provided $t_{\rm Q} \gg \Gamma_{\rm HI}^{-1}$. Hence, under this assumption, the \ion{H}{1} fraction $x_{\rm HI}$ at every pixel along the skewer is set by 
\begin{equation}
x_{\rm HI} = n_{\rm e} \alpha_{\rm A}^{\left[ {\rm HI}\right]} \left( n_{\rm e}\alpha_{\rm A}^{\left[ {\rm HI}\right]} + \Gamma_{\rm HI}^{\rm tot} \right)^{-1},
\end{equation}
where $\alpha_{\rm A}^{\left[ {\rm HI}\right]}$ is the \ion{H}{1} Case A recombination coefficient. We note that in our radiative transfer calculations we have assumed that all neutral helium, i.e., \ion{He}{1}, has been already singly ionized together with \ion{H}{1} at $z \gtrsim 6$. Hence, the \ion{He}{1} fraction is $x_{\rm HeI} \ll 1$. Therefore, the \ion{He}{1} electrons are taken into account when computing the fractions of \ion{He}{2} and \ion{H}{1}, whereas \ion{He}{2} electrons are used according to the computed $x_{\rm HeII}$ from the radiative transfer algorithm.

\subsection{Evolution of the Temperature}
\label{sec:temp_evol}

Consider an element of ideal gas in the expanding Universe exposed to the quasar radiation. The temperature $T$ of the gas element then evolves with time $t$ according to the first law of thermodynamics (see \citealp{Hui1997})
\begin{equation}
	\frac{{\rm d}T}{{\rm d}t} = -2HT + \frac{2T}{3\left(1 + \delta \right)}\frac{{\rm d}\delta}{{\rm d}t} - \frac{T}{\Sigma_{i}X_{i}}\frac{{\rm d}\Sigma_{i}X_{i}}{{\rm d}t} + \frac{2}{3k_{\rm B}n_{\rm b}}\frac{{\rm d}E}{{\rm d}t},
\label{eqn:temp_evol}
\end{equation} 
where $H$ is the Hubble parameter, $E$ is the internal energy of the gas, $n_{\rm b}$ is the number density of the baryonic particles (including electrons), and $k_{\rm B}$ is the Boltzmann constant. 

The first term on the right-hand side of eqn.~(\ref{eqn:temp_evol}) corresponds to adiabatic cooling due to the expansion of the Universe. The expansion cooling timescale of the low density photoionized gas in the IGM ($\Delta \lesssim 10$) is the Hubble time $t_{\rm H}$ \citep{Miralda1994}. Thus, because $t_{\rm H}\left(z \simeq 4 \right) \sim 2 \times 10^9$~yr is much longer than the quasar lifetimes we consider in this paper ($10^6\ {\rm yr} \leq t_{\rm Q} \leq 10^8\ {\rm yr}$), this term would result in a small change in the temperature of the gas over this timescale and therefore is neglected in our analysis. The second term describes the cooling or heating from structure formation, which also has a characteristic time of the Hubble time, $t_{\rm H}$. Thus, this term can be neglected for the same reason as the first one, i.e. $t_{\rm Q}/t_{\rm H} \ll 1$. The third term, which represents the change in the internal energy of the gas particles due to the change in the number of particles is also smaller than the dominant heating mechanism during reionization described by the last term ($\approx 4\%$; \citealp{Upton2016}), which accounts for the amount of heat gained (lost) per unit volume from radiation processes. Hence, assuming that hydrogen is already ionized and helium is singly ionized (\ion{He}{1} $\longrightarrow$ \ion{He}{2}), and that gas is exposed to the quasar radiation for the time $t_{\rm Q} \ll t_{\rm H}$, we can simplify eqn.~(\ref{eqn:temp_evol}) to
\begin{equation}
	\frac{{\rm d}T}{{\rm d}t} \simeq \frac{2}{3k_{\rm B}n_{\rm b}}\frac{{\rm d}E}{{\rm d}t} = \frac{2}{3k_{\rm B}n_{\rm b}} n_{\rm HeII} \epsilon_{\rm HeII},
\label{eqn:temp_evol2}
\end{equation} 
where $n_{\rm HeII}$ is the number density of \ion{He}{2} atoms and $\epsilon_{\rm HeII}$ is the total quasar photoheating rate. Using eqn.~(\ref{eqn:dxHeIIdt}) and given that the recombination timescale is very long compared to the photoionization timescale ($t_{\rm rec} \simeq 10^9~{\rm yr} \gg 1/\Gamma^{\rm tot}_{\rm HeII}$), and ignoring collisional ionization of \ion{He}{2}, the rate of change of \ion{He}{2} number density $n_{\rm HeII}$ is
\begin{equation}
	\frac{{\rm d} n_{\rm HeII}}{{\rm d} t} = - n_{\rm HeII} \Gamma_{\rm HeII}^{\rm tot}
\label{eqn:temp_evol2.1}
\end{equation} 
Therefore, combining eqn.~(\ref{eqn:temp_evol2}) and eqn.~(\ref{eqn:temp_evol2.1}) yields
\begin{equation}
	\frac{{\rm d}T}{{\rm d}t} \simeq \frac{2}{3k_{\rm B}n_{\rm b}} \biggl| \frac{{\rm d}n_{\rm HeII}}{{\rm d}t} \biggr| \frac{\epsilon_{\rm HeII}}{\Gamma_{\rm HeII}^{\rm tot}}.
\label{eqn:temp_evol3}
\end{equation} 
We can also write ${\rm d}n_{\rm HeII} / {\rm dt}$ in terms of \ion{He}{2} fraction $x_{\rm HeII}$ as
\begin{equation}
	\frac{{\rm d}n_{\rm HeII}}{{\rm d}t} = \frac{\rho_{\rm crit} \Omega_{\rm b}}{m_{\rm p}} \frac{Y_{\rm He}}{4} \left( 1 + \delta\right) \left( 1 + z\right)^3 \frac{{\rm d}x_{\rm HeII}}{{\rm d}t},
\label{eqn:temp_evol_xHeII}
\end{equation} 
where $\rho_{\rm crit}$ is the critical density of the Universe, $m_{\rm p}$ is the mass of proton, and $\left( 1 +\delta \right)$ is the gas density in units of the cosmic mean density. Finally, combining eqn.~(\ref{eqn:temp_evol3}) and eqn.~(\ref{eqn:temp_evol_xHeII}) and taking into account that the total number density of baryons is $n_{b} \approx n_{\rm H} + n_{\rm He} + n_{\rm e}$ (number density of hydrogen and helium atoms, and electrons) the temperature evolution is given by
\begin{equation}
	\frac{{\rm d}T}{{\rm d}t} \simeq \frac{2}{3k_{\rm B}}\frac{Y_{\rm He}}{4\left(2-5Y_{\rm He}/4\right)} \frac{{\rm d}x_{\rm HeII}}{{\rm d}t} \langle E \rangle
\label{eqn:temp_evol_final}
\end{equation}
where $\langle E \rangle = \epsilon_{\rm HeII} / \Gamma_{\rm HeII}^{\rm tot}$ is the average excess energy per per photoionization of a \ion{He}{2} atom given by 
\begin{eqnarray}\label{eqn:E}
\langle E \rangle = \Bigg[ \int_{\nu_{\rm th}}^{\infty} N_{\nu}\sigma_{\nu}e^{-\langle\tau_{\rm \nu}\rangle} \left(1-e^{-\langle\delta\tau_{\nu}\rangle}\right)  \left( h\nu -  h\nu_{\rm th}\right)& {\rm d}\nu \Bigg] \nonumber \\ \Bigg/ \Bigg[ \int_{\nu_{\rm th}}^{\infty}N_{\nu}\sigma_{\nu} e^{-\langle\tau_{\rm \nu}\rangle}\left(1-e^{-\langle\delta\tau_{\nu}\rangle}\right) {\rm d}\nu \Bigg]
\end{eqnarray}
We neglect the heating by the \ion{H}{1} and \ion{He}{2} ionizing
background, since this is accounted for via the treatment of
photoionization heating in the hydrodynamical simulation. Given the quasar SED slope $\alpha$ (eqn.~\ref{eqn:lum}), the frequency integral for the the average
excess energy per ionization $\langle E \rangle$ in eqn.~(\ref{eqn:E}) has an analytical
solution, which is then evaluated at each time step of our $1$D
radiative transfer calculations using the values of $\langle
\tau_\nu\rangle$ and $\langle \delta \tau_\nu\rangle$ computed by the
code. If we assume an optically thin limit and use eqn.~(\ref{eqn:temp_evol_final}), we can estimate the lower limit on expected heating produced by the quasar $\Delta T \simeq 0.6 h\nu_{\rm th} / [28\ k_{\rm B} \left( \alpha + 2\right)] \simeq 4500^{\circ}{\rm K}$ \citep{McQuinn2009}. In reality, however, the exact amount of heating depends on how many of hard photons are absorbed close to the quasar (i.e., their mean-free path), the properties of quasar itself and the history of \ion{He}{2} reionization.

\subsection{Examples Outputs from the Radiative Transfer Code}

\begin{figure}[!t]
\centering
 \includegraphics[width=1.0\linewidth]{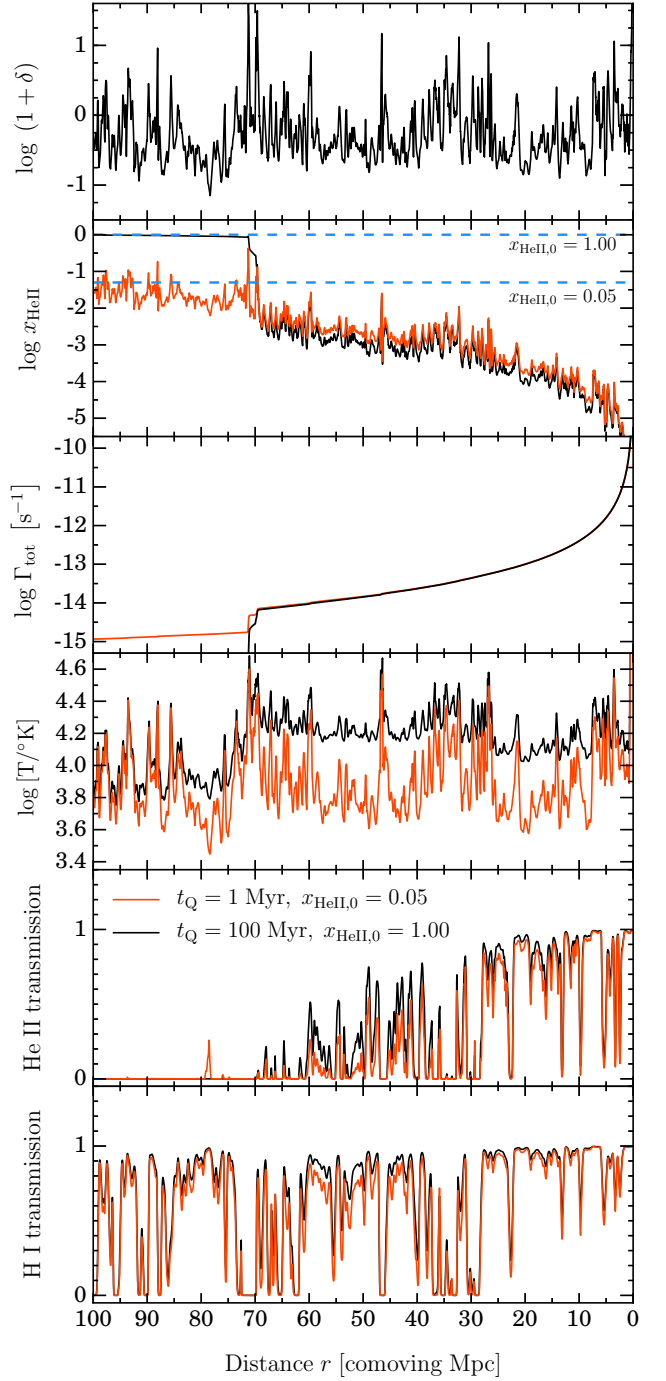}
 \caption{Example sightline at $z = 3.9$ from our radiative transfer calculations assuming a quasar turns on for $t_{\rm Q} = 10^6$~yr in IGM with initial \ion{He}{2} fraction $x_{\rm HeII,0} = 0.05$ (\emph{red}) and $t_{\rm Q} = 10^8$~yr in IGM with $x_{\rm HeII,0} = 1.00$ (\emph{black}). We indicate the initial \ion{He}{2}  fractions $x_{\rm HeII,0}$ before the quasar is on by the dashed lines in the panel with $x_{\rm HeII}$ evolution. The x-axis indicates distance $r$ from the quasar in units of comoving Mpc. Panels show (from top to bottom): the overdensity, the \ion{He}{2} fraction $x_{\rm HeII}$, the \ion{He}{2} photoionization rate $\Gamma_{\rm HeII}^{\rm tot}$, the temperature of the gas $T$, transmission in \ion{He}{2} and \ion{H}{1}, respectively.}
 \label{fig:TPE_ex}
\end{figure}

Following the approach described in \citet{Theuns1998}, we calculate \ion{H}{1} and \ion{He}{2} spectra along each of the skewers taken from the SPH simulations at all considered redshifts. Figure~\ref{fig:TPE_ex} illustrates an example output of our radiative transfer calculations with multiple physical quantities along one skewer. The quasar is located at $r = 0$. We show results from two models: the red curves assume a quasar which shines for $t_{\rm Q} = 10^6$~yr in an
IGM where helium is already largely doubly ionized with $x_{\rm HeII,0} = 0.05$, where $x_{\rm HeII,0}$ represents the \ion{He}{2} fraction before the quasar turns on. Whereas for the black curves the quasar shines for a longer $t_{\rm Q} = 10^8$~yr in an initially singly ionized IGM (i.e. before the epoch of \ion{He}{2} reionization) with $x_{\rm HeII,0} = 1.0$. The quasar photon production rates at \ion{H}{1} and \ion{He}{2} ionization thresholds are set to our fiducial values (see Section~\ref{sec:rt_lum}).

The uppermost panel of Figure~\ref{fig:TPE_ex} plots the gas density along the skewer in units of the cosmic mean density. The second panel from the top illustrates the \ion{He}{2} fraction for the two models. The horizontal dashed lines indicate the initial \ion{He}{2} fraction set by \ion{He}{2} ionizing background only, which prevailed in the IGM prior to the quasar turning on (see \S~\ref{sec:He2f}). As expected, close to the quasar (at $r \lesssim 20$~cMpc) helium is highly doubly ionized ($x_{\rm HeII,0} \lesssim 10^{-3.5}$) in both models due to the quasar's radiation. However, because the quasar photoionization rate drops off approximately as $r^{-2}$, which is illustrated in the third panel from the top, at larger radii it becomes sufficiently small and no longer dominates
over the \ion{He}{2} background. Therefore, the $x_{\rm HeII}$ in both models asymptote to the initial values set by $\Gamma_{\rm HeII}^{\rm bkg}$, i.e., $x_{\rm HeII,0} = 0.05$ (\emph{red}) and $x_{\rm HeII,0} = 1.00$ (\emph{black}), respectively.

The fourth panel from the top illustrates the IGM temperature along the skewer. According to eqn.~(\ref{eqn:temp_evol_final}) the heat input due to quasar radiation is proportional to the change in the \ion{He}{2} fraction as the quasar ionization front traverses through the surrounding IGM. Hence, the temperature in the $x_{\rm HeII,0} = 1.00$ model, where the quasar radiation significantly changes $x_{\rm HeII}$, is $\approx 2$ times higher, than that of the $x_{\rm HeII,0} = 0.05$ model, in which the IGM was already highly doubly ionized to begin with.

As expected, the \ion{He}{2} transmission (fourth panel from the top), which depends on $x_{\rm HeII}$, follows the general radial trend set
by the evolution of the \ion{He}{2} fraction. It is apparent from Figure~\ref{fig:TPE_ex} that, the sizes of \ion{He}{2} proximity zones in two illustrated models are nearly identical, despite the different values of $t_{\rm Q}$ and $\Gamma_{\rm HeII}^{\rm bkg}$ used in these models. Indeed, as we showed in
\citet{Khrykin2016}, there is a significant degeneracy between $x_{\rm HeII}$ and quasar lifetime $t_{\rm Q}$ in setting the sizes of \ion{He}{2} Ly$\alpha$
proximity zones. Remember that acquiring far ultraviolet \ion{He}{2} Ly$\alpha$ spectra of quasars at such high redshifts ($z \gtrsim 3.9$) is extremely  difficult, therefore, it will be hard to break this degeneracy and determine $t_{\rm Q}$ and $x_{\rm HeII}$ directly from the spectra. The apparent difference in IGM temperatures in two considered models is also difficult to distinguish from the effect of this degeneracy.

The \ion{H}{1} transmission, on the other hand, which can be directly probed with optical spectra 
to much higher redshifts $z \lesssim 6$, when hydrogen is highly ionized after the end of \ion{H}{1} reionization, is also sensitive to the heating of the intergalactic gas. The bottom panel of Figure~\ref{fig:TPE_ex} shows that \ion{H}{1} Ly$\alpha$ transmission in the hotter model (\emph{black} curve) with $x_{\rm HeII,0} = 1.00$ is smoother and shows less small-scale structure, than that of cooler $x_{\rm HeII,0} = 0.05$ model. This so-called ``thermal proximity effect" due to the nearby quasar radiation results from thermal Doppler broadening and the differences in IGM temperature between the two models (see middle panel of Figure~\ref{fig:TPE_ex}). In \S~\ref{sec:powspec} we show that this effect can be used to constrain the parameters of interest, i.e., quasar lifetime $t_{\rm Q}$ and the initial \ion{He}{2} fraction $x_{\rm HeII,0}$. But first, to build intuition about the physical mechanisms responsible for the shape of the IGM temperature profiles in the vicinity of quasars, we explore how these profiles depend on quasar lifetime and initial \ion{He}{2} fraction in the next section. 

\section{The Structure of Quasar Thermal Proximity Zones}
\label{sec:T_IGM}

\subsection{Temperature profiles: effects of $t_{\rm Q}$ and $x_{\rm HeII,0}$}
\label{sec:temp_prof}

\begin{figure}[t]
\centering
 \includegraphics[width=1.0\linewidth]{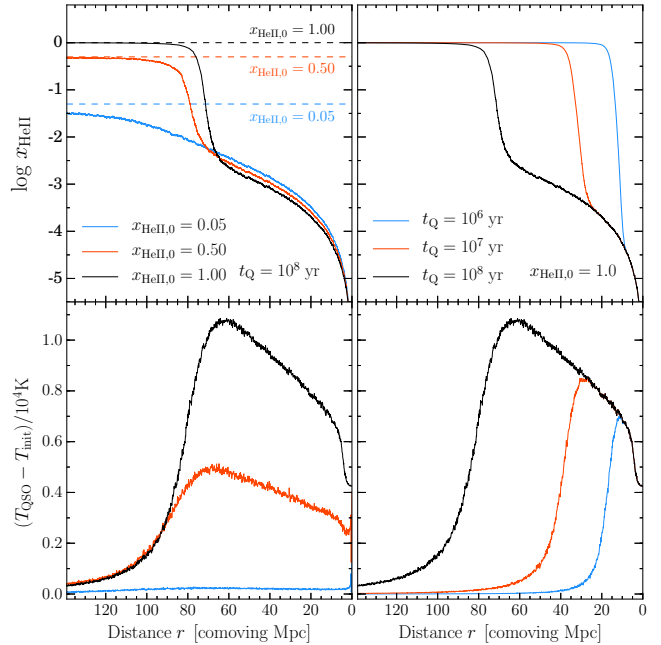}
 \caption{Thermal evolution of the intergalactic medium around the quasar in different radiative transfer simulations. The upper panels show the evolution of the median fraction of singly ionized helium $x_{\rm HeII}$, while the bottom panels illustrate the evolution of the median temperature boost in the IGM. \emph{Left side panels} show thermal evolution of the IGM as a function of initial \ion{He}{2} fraction at a fixed quasar lifetime $t_{\rm Q} = 10^8$~yr. Whilst \emph{right side panels} show thermal evolution of the IGM as a function of different values of quasar lifetime while keeping fixed initial \ion{He}{2} fraction $x_{\rm HeII,0} = 1.0$. The dashed lines in the upper left panel indicate the values of initial \ion{He}{2} fraction $x_{\rm HeII,0}$ in three different models. All median profiles are computed from $1000$ simulated skewers.}
 \label{fig:TPE_2}
\end{figure}

We calculate the median temperature and $x_{\rm HeII}$ profiles for several models using radiative transfer calculations for $1000$ skewers. The results are shown in Figure~\ref{fig:TPE_2}: the upper panels show the evolution of the median \ion{He}{2} fraction $x_{\rm HeII}$ and bottom panels show the median difference in temperature of the IGM $\Delta T$ prior to quasar turning on ($T_{\rm init}$) and after quasar activity ($T_{\rm QSO}$) for several different models\footnote{We compute the median profiles for illustration purposes only, because the mean profiles are much noisier due to sightline-to-sightline fluctuations and the effects of \ion{He}{2} LLSs.}. The panels at left show the evolution of these parameters as a function of initial \ion{He}{2} fraction $x_{\rm HeII,0}$, which prevailed in the IGM before the quasar turned on, while keeping the quasar lifetime fixed to $t_{\rm Q} = 10^8$~yr. Whereas, the panels on the right show models with three different values of quasar lifetime $t_{\rm Q}$ and the initial fraction of singly ionized helium fixed to $x_{\rm HeII,0} = 1.0$. Several trends are immediately
apparent from Figure~\ref{fig:TPE_2}.

\begin{figure*}[t]
\centering
 \includegraphics[width=0.8\linewidth]{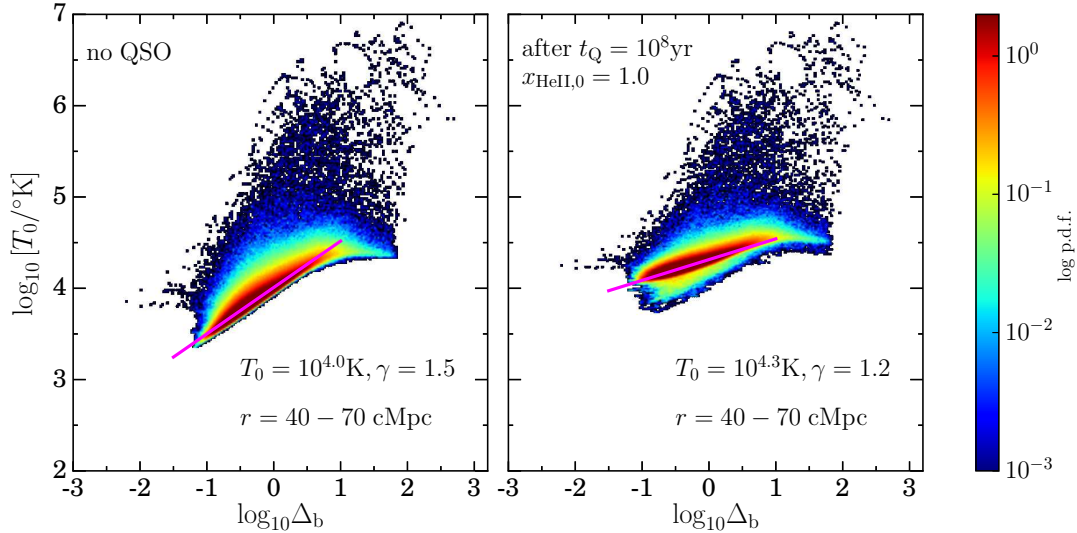}
 \caption{Temperature-density relation in simulations. \emph{Left:} prior to quasar turning on, \emph{right:} in radiative transfer simulations after quasar has been on for $t_{\rm Q} = 10^8$~yr in IGM with initial \ion{He}{2} fraction $x_{\rm HeII,0} = 1.00$. The relation is calculated at radial distances $r = 40 - 70$~cMpc where the temperature boost is maximal in the models we consider (see Figure~\ref{fig:TPE_2}).}
 \label{fig:TPE_3}
\end{figure*}

First, in \citet{Khrykin2016} we showed that the size of the \ion{He}{2} proximity zone, where the \ion{He}{2} fraction is greatly reduced by the quasar
ionization front sweeping across the IGM, depends strongly on the quasar lifetime $t_{\rm Q}$. This is clearly shown in the upper right panel of Figure~\ref{fig:TPE_2}. The same lifetime dependence is manifest in the temperature profiles in the bottom right panel of Figure~\ref{fig:TPE_2}. This is because for longer lifetimes (e.g., $t_{\rm Q} = 10^8$~yr) the quasar ionization front travels much further into the IGM, boosting the IGM temperature by $\Delta T \simeq 10^4$~K (in case of $x_{\rm HeII,0} = 1.0$). In contrast, the quasar ionization front has not yet reached the same distance in the short lifetime model (e.g., $t_{\rm Q} = 10^6$~yr), for which the size of the region is $\approx 4$ times smaller. In other words, the longer the IGM has been exposed to the quasar radiation, the larger the radial extent of the thermal proximity effect.

The panels at left in Figure~\ref{fig:TPE_2} indicate that the IGM temperature in the quasar proximity zone also depends significantly on the amount of initially singly ionized helium, $x_{\rm HeII,0}$. It is apparent that if $x_{\rm HeII,0} = 1.0$ (black curve), more hard photons emitted by the quasar photoheat the IGM and the median temperature is boosted by $\Delta T \simeq 10^4$~K, whereas there is no significant change in temperature if helium was already highly doubly ionized ($x_{\rm HeII,0} = 0.05$) before the quasar turned on (blue curve). According to eqn.~(\ref{eqn:temp_evol_final}) $\Delta T \sim {\rm constant}\times \Delta x_{\rm HeII}\langle E \rangle$, hence, if $x_{\rm HeII} \simeq 0.05$, then the temperature boost will be much smaller. To summarize, whereas the quasar lifetime determines the size of the thermal proximity zone, the value of initial \ion{He}{2} fraction sets the amplitude of the temperature boost in the zone.

Finally, it is apparent from bottom panels of Figure~\ref{fig:TPE_2}
that for fixed values of initial \ion{He}{2} fraction and quasar
lifetime the boost of IGM temperature is higher at larger distances
from the quasar. This is due to the filtering of the intrinsic quasar
spectrum by the IGM, which hardens the ionizing radiation field
\citep{Abel1999, McQuinn2009, Bolton2009, Meiksin2010}. Eqn.~(\ref{eqn:temp_evol_final}) indicates that the
amount of heat input over the time-step $t$ is proportional to the
average excess energy $\langle E \rangle$. Since the mean free path of
\ion{He}{2} ionizing photons scales as $\lambda_{\rm mfp} \propto
\nu^{3}$, photons with energy near the edge $E_{\rm ion} \simeq h\nu
\simeq 4$~Ry have a short mean free path. They, therefore, are
preferably absorbed by the IGM very near the quasar
resulting in a small value of $\langle E \rangle$. On the other hand, the high-energy photons can travel much
further into the IGM before getting absorbed, owing to their sufficiently
longer mean free path. They, thus inject more heat at larger
distances from the quasar.

\subsection{Temperature-Density Relation}
\label{sec:T-rho}

Eqn.~(\ref{eqn:temp_evol}) indicates that the temperature of the
intergalactic gas is determined by the competition between heating and
cooling processes, dominated by the photoheating due to metagalactic UV
background radiation and adiabatic cooling due to the expansion of the
Universe.\footnote{Gadget-3 code used in this study also models radiative cooling. It uses a cooling curve for primordial gas assuming an ionizing background with $\Gamma_{\rm HI}=10^{-12}{\rm s^{-1}}$ which takes a spectral index in the specific intensity $I_{\nu}$ of 0 \citep{Noh2014}.} This competition results in a tight relation between the temperature and density of the (unshocked) gas which can be approximated by a power-law $T = T_0 \Delta_{\rm b}^{\gamma - 1}$ (for $\Delta_{\rm b} \leq 10$), where $T_0$ is the temperature at mean density, $\gamma$ is the power-law
index, and  $\Delta_{\rm b} = \rho \slash \bar{\rho}$ is the overdensity. This power-law index asymptotes to a value $\gamma \approx 1.6$
\citep{Miralda1994, Hui1997, Hui2003, McQuinn2009, McQuinn2016} several hundred Myr after reionization events.

It is expected that \ion{He}{2} reionization will significantly change 
the temperature-density relation, tending to make it more isothermal, with $\gamma \longrightarrow 1$ \citep{McQuinn2009, Compostella2013, LaPlante2016}. This is
illustrated in Figure~\ref{fig:TPE_3} where we compare the
temperature-density relation from $1000$ skewers before quasar turns on
(left panel) with their values after these skewers have been
post-processed with our radiative transfer code, for a model with
$t_{\rm Q} = 10^8$~yr and $x_{\rm HeII,0} = 1.00$ (right panel). The
figure shows the temperature-density relation
at distances $r = 40 - 70$~cMpc from the
quasar, where according to Figure~\ref{fig:TPE_2} the temperature
boost is maximal for the models we consider. 
Whereas the original hydrodynamical simulation resulted in
temperature-density relation of ($T_0 = 10^4\,{\rm K}$, $\gamma = 1.5$), the
roughly density independent boost of $\Delta T \simeq 0.3$~dex in the
proximity zone results in ($T_0 = 10^{4.3}\,{\rm K}$, $\gamma = 1.2$).
Similar effects are seen in full $3$D radiative transfer simulations of \ion{He}{2} reionization \citep{McQuinn2009, Compostella2013, LaPlante2016}.

\begin{figure}[!t]
\centering
 \includegraphics[width=1.0\linewidth]{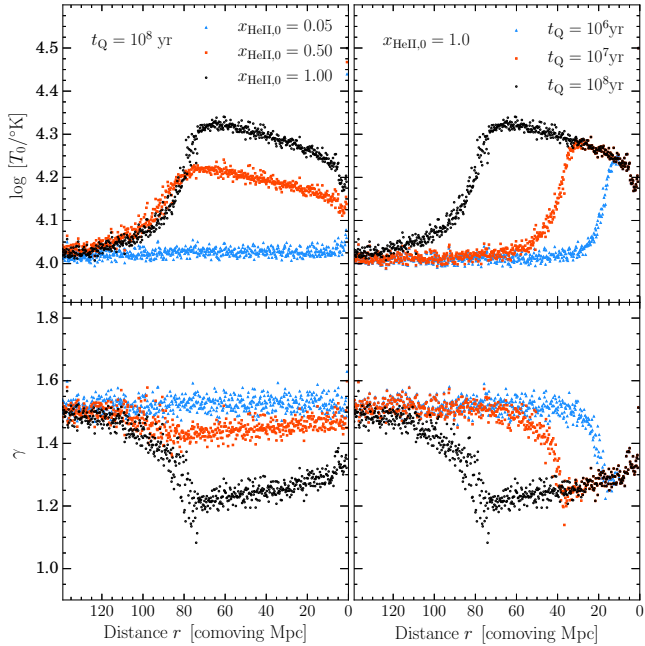}
 \caption[Radial evolution of the parameters of the temperature-density relation]{Parameters of the temperature-density relation: temperature at mean density $T_0$ and the power-law index $\gamma$, for the same set of radiative transfer simulations as in Figure~\ref{fig:TPE_2}.}
 \label{fig:TPE_4}
\end{figure}

In order to study how the quasar radiation impacts the
temperature-density relation with respect to our parameters of
interest $t_{\rm Q}$ and $x_{\rm HeII,0}$ we fit the distribution of
densities and temperatures in each pixel along $1000$ sightlines of
models shown in Figure~\ref{fig:TPE_2} with a power-law to calculate
$T_0$ and $\gamma$. To reduce the scatter due to density fluctuations,
we calculate the mean $T_0$ and $\gamma$ in bins of $20$ pixels, which
corresponds to $\Delta r \simeq 0.25$~cMpc. The results are shown in
Figure~\ref{fig:TPE_4}. As expected, the radial
profile of $T_0$ and $\gamma$ closely follows that of the IGM temperature boost, 
which in turn reflects the dependencies on \ion{He}{2} fraction and
quasar lifetime described previously (see Figure~\ref{fig:TPE_2}).

The main effect of quasar radiation on the temperature-density
relation is twofold. First, the intergalactic
medium becomes much hotter due to additional heat injected by the
quasar ionization front traversing through the IGM.
This is reflected by the increase in $T_0$ by $\Delta T_0 \simeq
0.3-0.4$ for $x_{\rm HeII,0} = 1.00$ model and smaller temperature
boosts for smaller \ion{He}{2} fractions. Second, the temperature-density relation flattens in the quasar
proximity zone with $\gamma$ deviating from the
asymptotic value $\gamma = 1.5$ (which prevailed in the IGM prior to
quasar turning on; see left panel of Figure~\ref{fig:TPE_3}) towards
the isothermal value of $\gamma = 1$. As illustrated previously, the
exact amplitude and extent of these changes strongly depend on the
value of the initial \ion{He}{2} fraction and quasar lifetime, as well as distance from the quasar in the thermal proximity zone. 
These trends are illustrated by the different curves in Figure~\ref{fig:TPE_4}.

To summarize, we investigated the dependence of the radial IGM
temperature profile around quasars on the initial \ion{He}{2} fraction
$x_{\rm HeII,0}$ and the quasar lifetime $t_{\rm Q}$. We demonstrated
that the radial extent of the elevated IGM temperatures probes the
quasar lifetime, whereas the amplitude of the temperature boost is set
by the amount of singly ionized helium in the IGM prior to the quasar
turning on. Therefore the thermal proximity effect can be used to
constrain both of these parameters and ultimately determine how quasar
lifetime evolves with redshift, and the redshift at which \ion{He}{2}
reionization occurred. In what follows we describe a method to detect
the thermal proximity effect and constrain these parameters. 

\section{Line-of-sight power spectrum statistics}
\label{sec:powspec}

Among the many methods used to study thermal evolution of the IGM the
Ly$\alpha$ forest flux power spectrum is the most sensitive probe of
the temperature of the gas in the IGM. This is due to the fact that thermal Doppler broadening affects the
properties of the \ion{H}{1} Ly$\alpha$ forest resulting in a prominent small-scale (high-$k$) cut-off in the
power \citep{McDonald2000, Zaldarriaga2001, Croft2002, Viel2009}. In this section we show that by measuring the power spectrum in bins of radial distance from the quasar, one can detect the thermal proximity effect and constrain the \ion{He}{2} fraction $x_{\rm HeII,0}$ and quasar lifetime $t_{\rm Q}$. In \S~\ref{sec:wavelets} we also discuss another method to detect the thermal proximity effect based on wavelet analysis.

Recall Figure~\ref{fig:TPE_2} where we showed that the amplitude and
the extent of the thermal proximity effect around the quasar have a
strong radial dependence which is sensitive to the quasar lifetime,
and a temperature boost sensitive to the initial singly ionized
fraction. Hence, the thermal broadening of the \ion{H}{1}
Ly$\alpha$ forest will be a function of quasar lifetime $t_{\rm Q}$,
initial \ion{He}{2} fraction $x_{\rm HeII,0}$ and distance $r$ from
the quasar. Our approach is therefore to calculate the average
\ion{H}{1} Ly$\alpha$ forest power spectrum of a sample of quasars (we
use a total number of $N = 1000$ simulated \ion{H}{1} Ly$\alpha$
spectra) in bins of $\Delta r = 10$~cMpc and to compare the results
with the power spectra from control regions far away from quasars and
hence outside of their proximity zones. Very close to the quasar
strong absorbers intrinsic to the quasar environment and shock-heated
gas will complicate our analysis, and this galaxy formation physics
will not be properly captured by our hydrodynamical simulations, hence
we will exclude the first $3$~cMpc closest to the quasar along each line-of-sight.

The flux contrast along the line-of-sight can be written as
\begin{equation}\label{eqn:flux_contr}
\delta F\left( x \right) = \frac{F \left( x \right) } {\langle F \rangle} - 1, 
\end{equation}
where $F \left( x \right) = {\rm exp}\left(-\tau\right)$ is the
transmitted flux in each pixel $x$ of the skewer with optical depth
$\tau$, and $\langle F \rangle$ is the mean flux of the IGM at any
given redshift, for which we adopt measurements of
\citet{Becker2013a}. The line-of-sight power spectrum of the Ly$\alpha$ forest $P\left( k|r \right)$ as a function of wavenumber $k$ and distance from the quasar $r$ is then given by
\begin{equation}\label{eqn:power}
P\left(k|r\right) = \langle \small| \delta \tilde{F}\left( k\right)\small|^2\rangle_N
\end{equation}
where $\delta \tilde{F}\left(k\right)$ corresponds to the Fourier
transform of $\delta F$ at wavenumber $k$, for a chunk of spectra in a
$\Delta r = 10$~cMpc at distance $r$ from the quasar, and angular
brackets denote the averaging over total ensemble of skewers $N$. We
adopt the common convention of working with the dimensionless power
spectrum $\pi^{-1}kP\left( k | r \right)$. Note, given the described binning in radial direction, we cannot measure $k$-modes larger than the fundamental mode of the $10$~cMpc radial bin, hence the range of $k$-modes we consider is $k = \left[ 0.007, 0.3 \right]\ {\rm km^{-1} s}$, where $k_{\rm min} = 2\pi \slash 10\ {\rm cMpc} \simeq 0.007\ {\rm km^{-1} s}$ and $k_{\rm max}$ is limited to $k_{\rm max} \simeq 0.3\ {\rm km^{-1}s}$. The range of $k$ modes between $k_{\rm min}$ and $k_{\rm max}$ is also divided into $10$ bins that are equally spaced in ${\rm log}\ k$ space.

In the following sections we investigate the sensitivity of this power spectrum computed in radial
bins to the value of the quasar lifetime, and the initial fraction of singly ionized helium.

\subsection{Sensitivity to Quasar Lifetime}
\label{sec:sql}

\begin{figure*}
\centering
 \includegraphics[width=1.\linewidth]{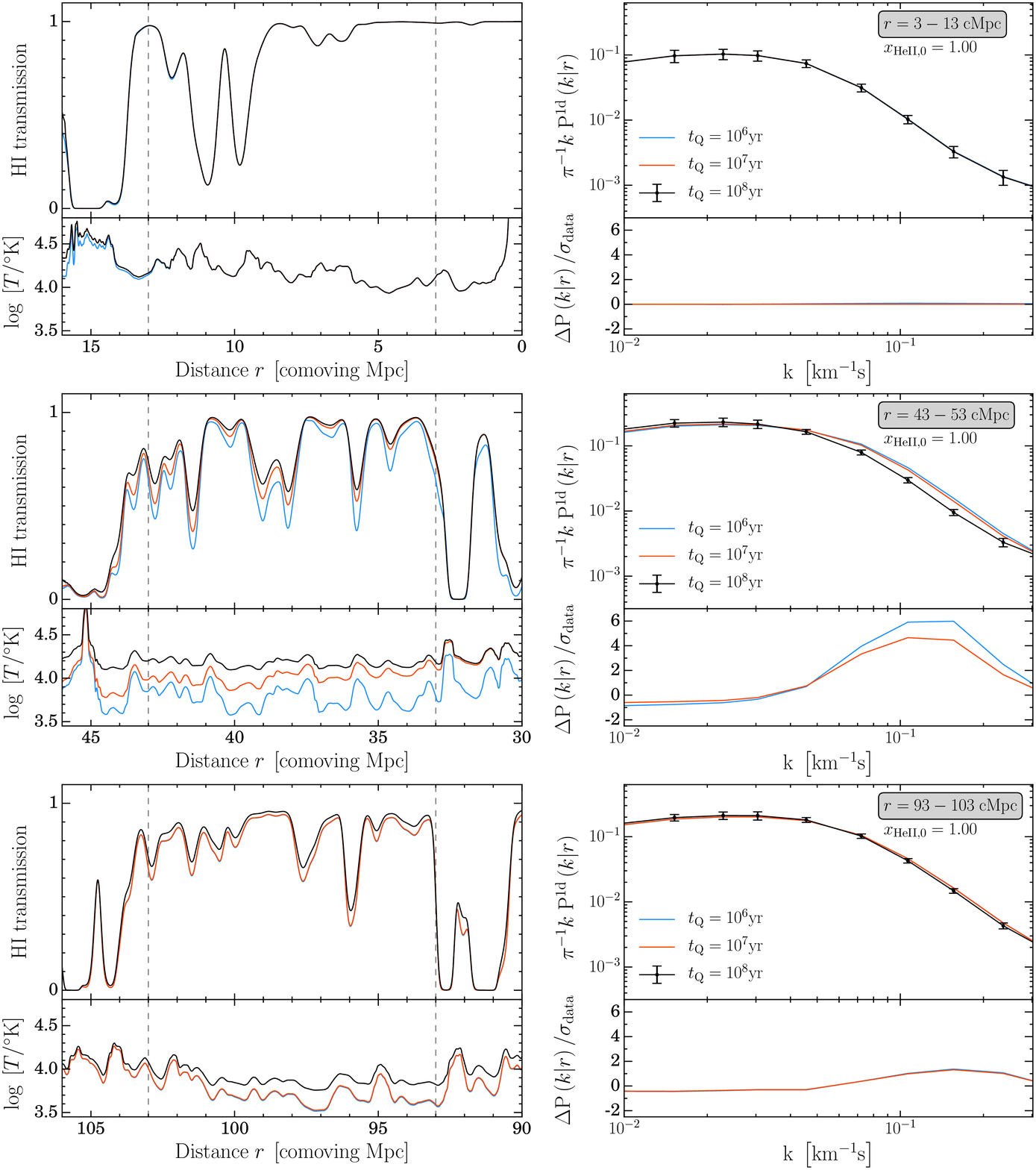}
 \caption[Sensitivity of the \ion{H}{1}\ Ly$\alpha$ forest power spectrum statistics to the value of quasar lifetime]{Sensitivity of the \ion{H}{1} Ly$\alpha$ forest power spectrum statistics to the value of quasar lifetime. Each row corresponds to the different radial bin from the quasar. Left column illustrates the simulated individual \ion{H}{1} Ly$\alpha$ spectra (top) and IGM temperature profiles (bottom) for three models with $t_{\rm Q} = 10^6$~yr (blue), $t_{\rm Q} = 10^7$~yr (red) and $t_{\rm Q} = 10^8$~yr (black). The value of initial fraction of singly ionized helium is fixed to $x_{\rm HeII,0} = 1.0$ in all models. Right column shows the average \ion{H}{1} Ly$\alpha$ power spectra (top subpanels) of the same models. The difference between the power spectrum of $t_{\rm Q} = 10^8$~yr model and the other models divided by the simulated error (see text for explanation) is shown in the bottom subpanels.}
\label{fig:PS_1}
\end{figure*}

We compute three different models with $t_{\rm Q} = 10^6$~yr, $t_{\rm
  Q} = 10^7$~yr, and $t_{\rm Q} = 10^8$~yr, while the value of the
initial \ion{He}{2} fraction is fixed to $x_{\rm HeII,0} = 1.0$. The
left columns of Figure~\ref{fig:PS_1} show simulated individual
\ion{H}{1} Ly$\alpha$ transmission spectra and temperature profiles in
the proximity zone for these three models, calculated in three
different radial bins, while the right column illustrates the
corresponding \ion{H}{1} Ly$\alpha$ power spectra averaged over $1000$
skewers in the same radial bins. The errorbars $\sigma_{\rm data}$, computed from $500$ random realizations of \emph{data} power spectrum ($N = 50$ skewers) drawn from original $1000$ skewers of the model with $t_{\rm Q} = 10^8$~yr and $x_{\rm HeII,0} = 1.00$, serve as an example of the constraining power
of a realistic dataset \footnote{The choice of $N=50$ skewers is based on the number of existing archival high-resolution \ion{H}{1}
  Ly$\alpha$ forest spectra covering $z\simeq 4$.}. The bottom subpanels in the right column also show the difference $\Delta P$ in
each radial bin between the power spectrum of each model and the power spectrum of our fiducial model (i.e., $t_{\rm Q} = 10^8$~yr and $x_{\rm HeII,0} = 1.00$)
divided by the simulated error $\sigma_{\rm data}$. There are several trends that are immediately apparent from Figure~\ref{fig:PS_1}.

First, in the bin closest to the quasar ($r = 3-13$~cMpc)
the time required by the ionization front to reach this distance and 
photoionize the gas is less than the quasar lifetimes for all three
of the models shown. Thus, very close to the quasar the amplitude of the temperature
boost is the same independent of the quasar lifetime (see
Figure~\ref{fig:TPE_2}), and the resulting \ion{H}{1} Ly$\alpha$
forest for these three lifetimes are indistinguishable in this bin,
and the average power spectra are identical. 

\begin{figure*}[!t]
\centering
 \includegraphics[width=1.\linewidth]{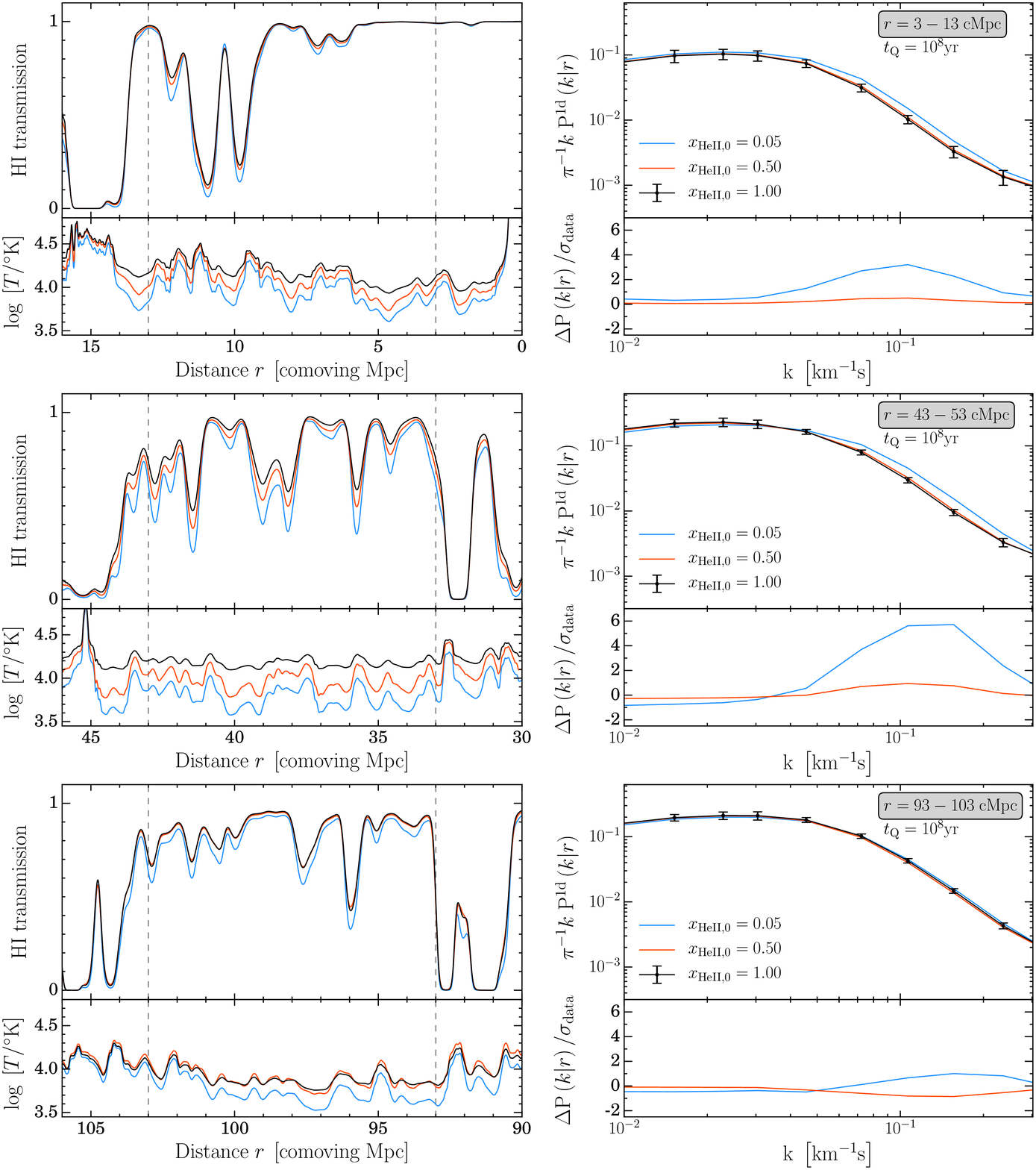}
 \caption[Sensitivity of the \ion{H}{1}\ Ly$\alpha$ forest power spectrum statistics to the value of initial \ion{He}{2}\ fraction]{Sensitivity of the \ion{H}{1} Ly$\alpha$ forest power spectrum statistics to the value of initial \ion{He}{2} fraction. Three models are plotted with initial \ion{He}{2} fraction $x_{\rm HeII,0} = 0.05$ (blue), $x_{\rm HeII,0} = 0.50$, and $x_{\rm HeII,0} = 1.00$. The lifetime of quasar is fixed in all models to $t_{\rm Q} = 10^8$~yr. See Figure~\ref{fig:PS_1} for the description of the panels.}
\label{fig:PS_2}
\end{figure*}

However, further away from the quasar the difference between power spectra increases.
It is apparent that at an intermediate distance $r = 43 - 53$~cMpc the
Ly$\alpha$ forest spectra and temperature profiles of the three models differ
significantly.  For the longest lifetime model the quasar ionization
front has already traversed and ionized gas at this distance and boosted the IGM
temperature by $\Delta T \approx 10^4$~K (see
Figure~\ref{fig:TPE_2}), whereas it has not yet reached this distance
in models with shorter quasar lifetime ($t_{\rm Q} = 10^7$~yr and
$t_{\rm Q} = 10^6$~yr). Thus, the temperature of this region of the IGM is
lower in these models, than in $t_{\rm Q} = 10^8$~yr model. This
difference in the IGM temperature produces different amounts of
thermal broadening in the Ly$\alpha$ forest, that is why the
\ion{H}{1} Ly$\alpha$ forest of the longest lifetime model becomes
smoother and contains less saturated absorption features, than
\ion{H}{1} Ly$\alpha$ spectra of cooler models. Accordingly, the power
spectrum of the $t_{\rm Q} = 10^8$~yr models exhibits less small-scale power (high-$k$),
because it is smoothed out by the thermal broadening. 
Thus, we conclude that there is considerably more small-scale power 
($k \gtrsim 0.05\ {\rm km^{-1} s}$) in models with shorter quasar lifetimes. The differences between the power spectra in
this radial bin, illustrated in the bottom subpanel, shows that
these models should be easily distinguishable.

Finally, far away from the quasar at $r \simeq 93-103$~cMpc the difference between the power spectra of $t_{\rm Q} = 10^6$~yr and $t_{\rm Q} = 10^7$~yr models diminishes again because there was not enough time for the quasar ionization front to reach and ionize gas at this distance, and significantly change the temperature of the IGM, whereas the thermal proximity zone for the $t_{\rm Q} = 10^8$~yr model extends out to $r \sim 100$~cMpc and still exhibits a small boost $\Delta T \approx 10^3$~K (see Figure~\ref{fig:TPE_2}). Hence the $t_{\rm Q} = 10^6$~yr and $t_{\rm Q} = 10^7$~yr models are indistinguishable in this bin and the gas is at the ambient IGM temperature, whereas these models differ from the $t_{\rm Q} = 10^8$~yr model. 

\subsection{Sensitivity to initial He II Fraction}

Similarly, we now investigate the sensitivity of the line-of-sight
Ly$\alpha$ power spectrum statistics to the variations of the initial
\ion{He}{2} fraction. Akin to the discussion in previous section we
run three radiative transfer simulations, each with different values
of the initial \ion{He}{2} fraction $x_{\rm HeII,0} = 0.05$, $x_{\rm
  HeII,0} = 0.50$, and $x_{\rm HeII,0} = 1.00$, respectively. The
quasar lifetime is fixed to $t_{\rm Q} =
10^8$~yr. Figure~\ref{fig:PS_2} illustrates the resulting \ion{H}{1}
Ly$\alpha$ transmission, IGM temperature profiles and power spectra of
these models in the same radial bins as Figure~\ref{fig:PS_1}.

In \S~\ref{sec:temp_evol} we showed that the initial \ion{He}{2} fraction determines the amplitude of the temperature boost in the proximity zone (see left panels of Figure~\ref{fig:TPE_2}). Thus, one naturally expects to see significant differences in the properties of the \ion{H}{1} Ly$\alpha$ forest and its power spectrum between models with very different values of $x_{\rm HeII,0}$. Indeed, one can see from the left column of Figure~\ref{fig:PS_2} that the temperature of the IGM significantly differs between the considered models (reaching maximum $\Delta T \approx 10^4$~K at $r \simeq 43 - 53$~cMpc). As expected, the temperature boost in the proximity zone is maximized in the $x_{\rm HeII,0} = 1.00$ model (black). On the contrary, as explained in Section~\ref{sec:temp_evol}, the temperature of the IGM hardly changes if helium is predominantly highly doubly ionized before the quasar turns on ($x_{\rm HeII,0} = 0.05$, blue curve). Analogous to the discussion in Section~\ref{sec:sql}, this difference in the photoheating results in different amount of thermal broadening of the \ion{H}{1} Ly$\alpha$ forest, therefore affecting the power spectra of the models, i.e. decreasing the small-scale power in the hotter model ($x_{\rm HeII,0} = 1.00$) in comparison to the cooler one ($x_{\rm HeII,0} = 0.05$).

Interestingly, the difference between the power spectra of the $x_{\rm
  HeII,0} = 1.00$ (black) and two models with $x_{\rm HeII,0} = 0.05$
(blue) and $x_{\rm HeII,0} = 0.50$ (red) first increases as one goes
from the bin closest to the quasar ($r = 3-13$~cMpc) to the
intermediate bin ($r = 43 - 53$~cMpc) and decreases afterward as one
goes to the outskirts of the proximity zone ($r = 93-103$~cMpc). This
behavior is driven by two effects. First, as we noted in
Section~\ref{sec:T_IGM}, the intergalactic medium filters quasar
radiation, which results in more heat injected at larger distances
from the quasar, than close to it (see
Figure~\ref{fig:TPE_2}). Consequently, the thermal broadening of the
\ion{H}{1} Ly$\alpha$ forest lines is strongest at the intermediate
distance $r = 43 - 53$~cMpc ($\Delta T \approx 10^4$~K).
Hence, the difference between the average power spectra illustrated in the right columns of
Figure~\ref{fig:PS_2} follows similar behavior. However, recall that
the radial extent of the boost of the IGM temperature depends on the
value of quasar lifetime, which is fixed to $t_{\rm Q} = 10^8$~yr in
all models considered here. As we showed in Section~\ref{sec:sql} the
temperature of the IGM is increased only at distances $r$ for which the quasar ionization front has had enough time to travel to and ionize the
gas. At further distances from the quasar, the temperature asymptotes to the
level of the ambient IGM (see right panels in
Figure~\ref{fig:TPE_2}). Therefore, at the largest radii the power spectra
of all three models approach each other and the differences are significantly
smaller. 

We have demonstrated the sensitivity of the \ion{H}{1} Ly$\alpha$
forest power spectrum to the parameters defining the radial extent and
the amplitude of the thermal proximity effect: quasar lifetime $t_{\rm
  Q}$ and the initial fraction of singly ionized helium $x_{\rm
  HeII,0}$.  We now proceed to estimate the accuracy with which
we can constrain these parameters by applying a Bayesian MCMC analysis
to a mock sample of \ion{H}{1} Ly$\alpha$ forest power spectra.

\section{Estimation of the quasar lifetime and \ion{He}{2} fraction}
\label{sec:est}

Our subsequent analysis of the \ion{H}{1} Ly$\alpha$ forest power
spectrum assumes that the temperature of the ambient IGM in regions
far away from active quasars has already been measured to high
precision. Hence, in what follows we imagine that the hydrodynamical
simulations that we use as the input to our radiative transfer
calculations are designed to reproduce these temperature measurements,
and thus to have the correct temperature-density relation in the IGM prior to when the quasars responsible for the thermal proximity effect
turned on. Although the ambient IGM's temperature-density relation may very well be determined by similar
thermal proximity effects from a previous generation of quasars, we
can nevertheless imagine that the quasars whose proximity zones we
study at any epoch turned on in an IGM whose thermal state is
precisely known. Assuming that the outputs from our hydrodynamical
simulations have been calibrated in this way, we then perform
radiative transfer calculations for different combinations of $t_{\rm
  Q}$ and $x_{\rm HeII,0}$. Given the dependence of the \ion{H}{1}
Ly$\alpha$ forest power spectrum on these parameters (see
Section~\ref{sec:powspec}), we compare the power spectra of different
models and therefore deduce the precision with which quasar lifetime
and initial \ion{He}{2} fraction can be determined. We make no attempt
to model or marginalize out uncertainties on the thermal state of the
ambient IGM. In what follows we describe the basic aspects of our
method and begin with the definition of the likelihood required by the MCMC algorithm.

\subsection{The Likelihood}
\label{sec:like}

\begin{figure*}[!t]
\centering
 \includegraphics[width=0.8\linewidth]{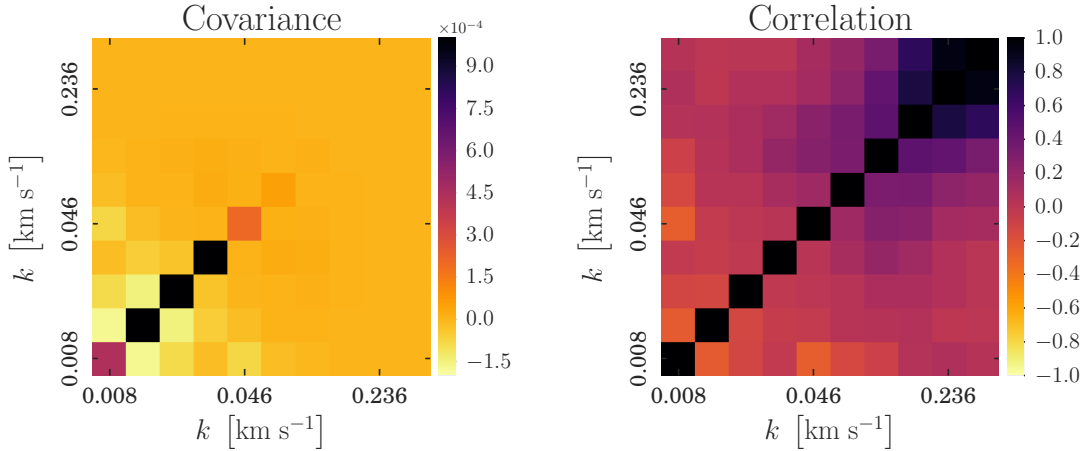}
 \caption{Example of the covariance and correlation matrices in a single radial bin $r = 63-73$~cMpc (see eqn.~\ref{eqn:cov}-\ref{eqn:cor}) for the \emph{data} sample is drawn from the model with $t_{\rm Q} = 10^8$~yr and $x_{\rm HeII,0} = 1.0$.}
\label{fig:Auto-cov}
\end{figure*}

We construct a grid of $81$ models at $z = 3.9$ ($1000$ skewers per model) with combination of initial \ion{He}{2} fraction and logarithmically spaced quasar lifetime values, where $x_{\rm HeII, 0} = \left[ 0.05, 0.10, 0.20, 0.30, 0.50, 0.60, 0.70, 0.90, 1.00 \right]$ and ${\rm log} \left( t_{\rm Q}/{\rm Myr}\right) = \left[ 0.0, 0.25, 0.5, 0.75, 1.0, 1.25, 1.5, 1.75, 2.0 \right]$.
Further, similar to the discussion in
\S~\ref{sec:powspec}, we choose the \emph{data} sample to be
represented by $N = 50$ \ion{H}{1} Ly$\alpha$ spectra. In what follows
we perform MCMC inference for $9$ different \emph{data} samples, each represented by
$50$ \ion{H}{1} Ly$\alpha$ spectra drawn from one model with one
combination of $\lbrace x_{\rm HeII,0}^{\rm data}, {\rm log}\ t_{\rm
  Q}^{\rm data} \rbrace$, where $x_{\rm HeII,0}^{\rm data} = \left[
  0.05, 0.5, 1.0\right]$ and ${\rm log} \left( t_{\rm Q}^{\rm
  data}/{\rm Myr}\right) = \left[ 0.0, 1.0, 2.0 \right]$.
It is well known that the distribution of power spectrum measurements
is well described by a multi-variate Gaussian distribution, and following the standard
approach \citep{McDonald2006,PD2013}, we adopt the following form for our likelihood of the \emph{data} power spectrum given the model in each
radial bin:
\begin{eqnarray}
\mathscr{L}^{\rm bin}\left\{ P_{\rm data}\left(k|r\right) | x_{\rm HeII,0}, {\rm log}\ t_{\rm Q}, r \right\} = \left( 2\pi \right)^{-{\rm N_{\rm k}}/2} \times \nonumber\\
 {\rm det}\left( \Sigma \right)^{-1/2} {\rm exp} \Bigg( - \frac{1}{2} \bigg[ P_{\rm data}\left(k|r\right) - P_{\rm model} \left(k|r\right)&\bigg]^{\rm T} \nonumber \\
\times \Sigma_{\rm bin}^{-1} \bigg[ P_{\rm data} \left(k|r\right) - P_{\rm model}\left(k|r\right) \bigg] \Bigg),\ 
\label{eqn:likelihood}
\end{eqnarray}
where $N_{\rm k}$ is the number of band powers that we consider in each radial bin, $P_{\rm data}\left(k|r\right)$ is the average power spectrum of $N=50$ \emph{data} skewers in each radial bin, $P_{\rm model}\left(k|r\right)$ is the average power spectrum of each model on the grid in $2$D parameter space in this radial bin (calculated from $N = 1000$ skewers), and $\Sigma_{\rm bin}$ is the covariance matrix of the \emph{data} power spectrum in the corresponding
radial bin. The covariance is given by 
\begin{eqnarray}
\Sigma_{\rm bin} \left(k, k^\prime \right) = \Bigg \langle \Big[ P_{\rm data}\left(k|r\right) - \langle P \left(k|r\right)\rangle \Big] \times \nonumber\\
\Big[ P_{\rm data}\left(k^\prime|r\right) - \langle P \left(k^\prime|r\right)\rangle \Big] \Big \rangle_{\rm N}
\label{eqn:cov}
\end{eqnarray}
where $\langle P\left(k|r\right) \rangle$ is the average power spectrum of $1000$ skewers drawn from the model with the same parameters as the \emph{data} subsample. For each of the 9 different models (different combinations of $\{ x_{\rm HeII,0}, {\rm log}\ t_{\rm Q}\} $)
for which we perform MCMC inference, the covariance matrix $\Sigma_{\rm bin}$ is
computed by calculating $P_{\rm data}$ from $N = 500$ random realizations of
$N=50$ \emph{data} skewers drawn from each model. An example of the covariance and correlation matrices given by eqn.~(\ref{eqn:cov}) is shown in
Figure~\ref{fig:Auto-cov} for a model with $t_{\rm Q} = 10^8$~yr and
$x_{\rm HeII,0} = 1.00$, where the correlation matrix illustrates the degree of correlation between pairs of pixels $\lbrace k, k^\prime \rbrace$ and defined as
\begin{equation}
R_{\rm bin}\left(k, k^{\prime} \right) =  \frac{\Sigma_{\rm bin}(k,k^{\prime})}{\sqrt{\Sigma_{\rm bin}(k,k)\Sigma_{\rm bin}(k,k^{\prime})}}
\label{eqn:cor}
\end{equation}

\begin{figure*}[!t]
\centering
 \includegraphics[width=1.0\linewidth]{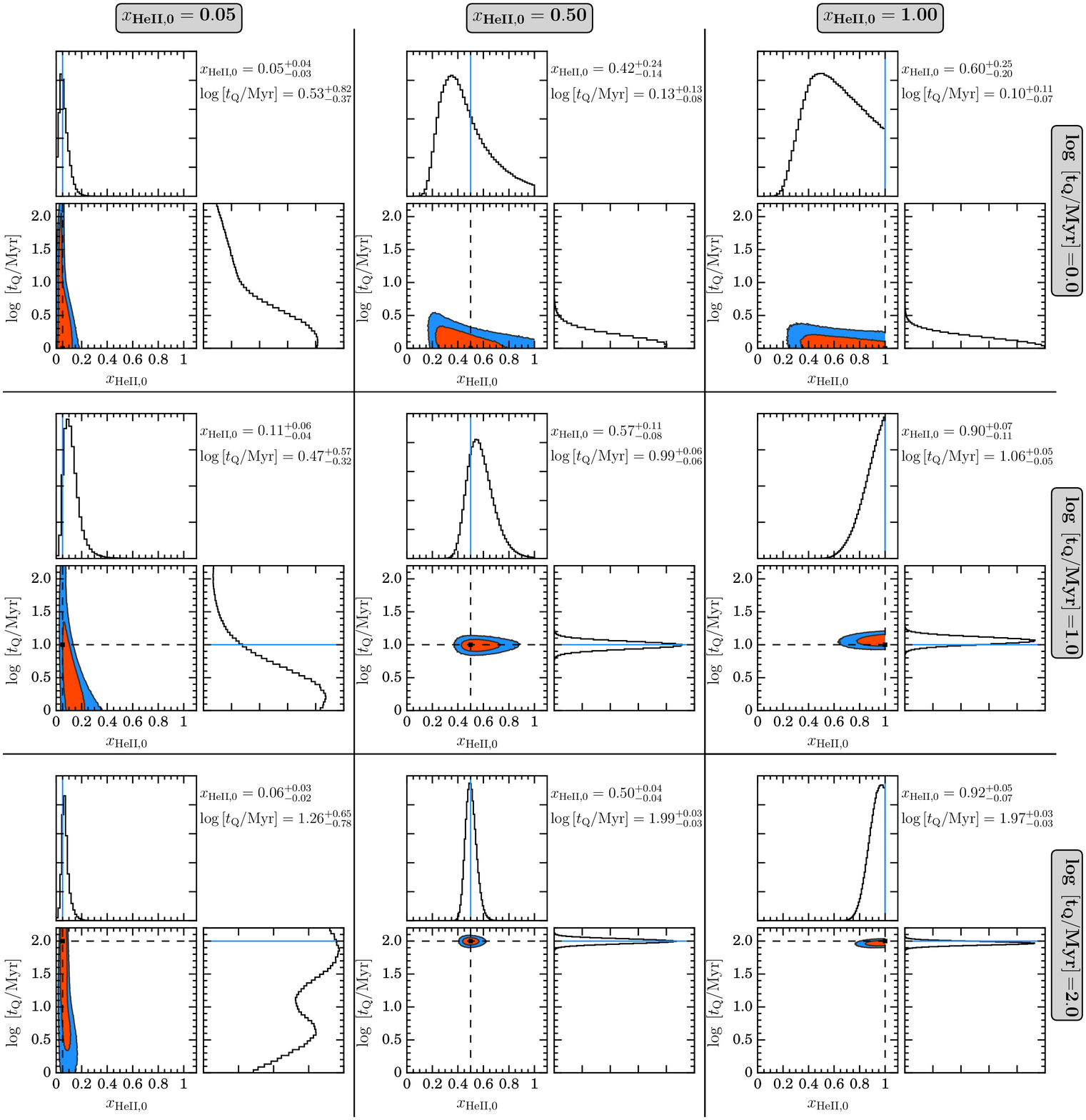}
 \caption{Constraints on quasar lifetime and initial \ion{He}{2} fraction from the MCMC analysis. The labels on top of each column and on the right side of each row show the values of initial \ion{He}{2} fraction and quasar lifetime in the sample of $N = 50$ skewers representing \emph{data}. The $95\%$ (red) and $68\%$ (blue) confidence levels from MCMC calculations are shown in each panel. The dashed lines show the values of lifetime and \ion{He}{2} fraction in the sample representing \emph{data}. The quoted values of $x_{\rm HeII,0}$ and $t_{\rm Q}$ are the $16$th, $50$th and $84$th percentiles of the corresponding marginalized distributions.}
\label{fig:MCMC}
\end{figure*}

\begin{figure}[b]
\centering
 \includegraphics[width=1.0\linewidth]{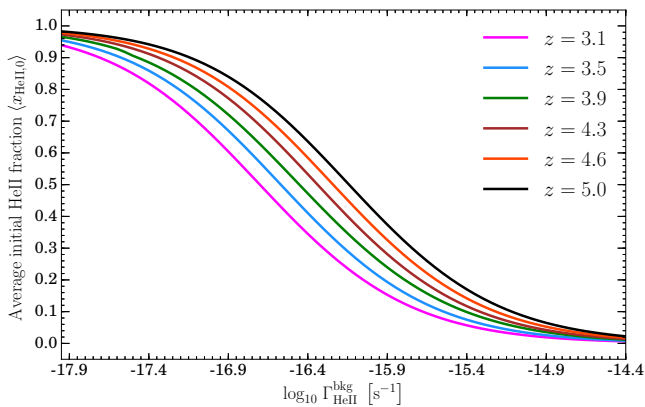}
 \caption{Average initial \ion{He}{2} fraction $\langle x_{\rm HeII,0} \rangle$ in our radiative transfer calculations as a function of the assumed \ion{He}{2} ionizing background (see Section~\ref{sec:model} for details) for each value of redshift we consider. The parameter grids at each redshift were constructed in accordance to this dependence.}
\label{fig:xHeII_Gamma}
\end{figure}

\begin{figure*}[t]
\centering
 \includegraphics[width=1.0\linewidth]{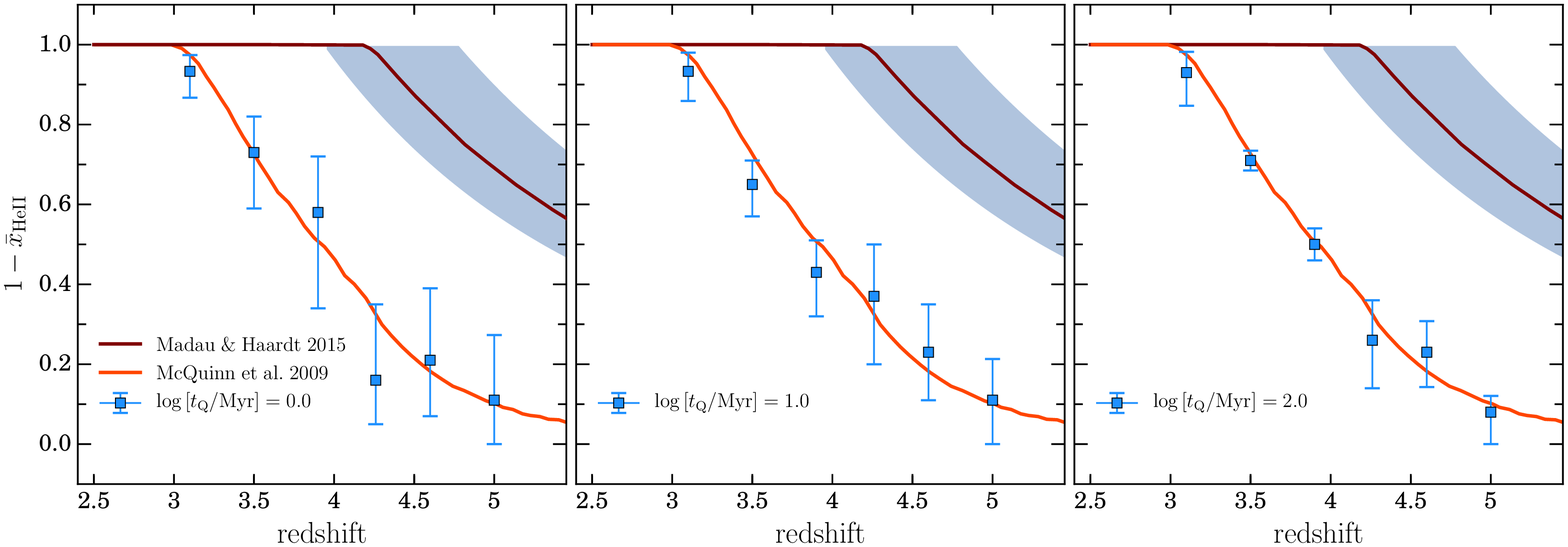}
 \caption{Reconstructed history of \ion{He}{2} reionization, quantified by the fraction of completely doubly ionized helium $x_{\rm HeIII} \equiv 1 - x_{\rm HeII}$. Each panel corresponds to assumed value of quasar lifetime in the \emph{data} samples, i.e., $t_{\rm Q} = 10^6$~yr (left), $t_{\rm Q} = 10^7$~yr (middle), and $t_{\rm Q} = 10^8$~yr (right), respectively. The solid lines in all three panels illustrate the results of the simulations of \ion{He}{2} reionization history: \emph{red} - \citet{McQuinn2009}, \emph{brown} - \citet{Madau2015} ($1\sigma$ variance in their models is illustrated by grey shaded area). The blue data points in each panel show the results of our calculations, specifically the $50th$ percentile from MCMC $1$D posterior distributions of $x_{\rm HeII,0}^{\rm data}$. The errorbars indicate $16{\rm th}$ and $84{\rm th}$ percentiles of the same distributions (see Section~\ref{sec:mcmc} for details), respectively. The number of spectra used in \emph{data} sample at redshifts $z = \left[ 3.1, 3.5, 3.9\right]$ is $N = 50$, while at $z = \left[ 4.3, 4.6, 5.0 \right]$ $N = 20$, based on the number of available to date quasar spectra.}
\label{fig:history}
\end{figure*}

Following eqn.~(\ref{eqn:likelihood}) we can compute the likelihood of the \emph{data} power spectrum in each radial bin inside the thermal proximity region for any given model.  The full likelihood of the \emph{data} in the
proximity zone is then calculated multiplying the corresponding likelihoods in each radial bin, yielding
\begin{equation}
\mathscr{L}^{\rm full} \left( P_{\rm data}\left(k|r\right) | x_{\rm HeII,0}, {\rm log}\ t_{\rm Q} \right) = \prod_{i=1}^{\rm M} \mathscr{L}^{\rm bin}_{i}
\label{eqn:likelihood_full}
\end{equation}
where ${\rm M} = 11$ is the number of $10\ {\rm cMpc}\ r$-bins used
(from $r = 3$~cMpc to $r =
113$~cMpc). Eqn.~(\ref{eqn:likelihood_full}) is valid under the
assumption that the correlations between the power spectra in the
neighboring radial bins are small or negligible. We justify this
assumption in Appendix~\ref{ap:inf_test}, to which we refer the
interested reader. We use eqn.~(\ref{eqn:likelihood_full}) to
calculate $\mathscr{L}^{\rm full}$ at each point in our parameter
grid, and then use bivariate spline interpolation to estimate
$\mathscr{L}^{\rm full}$ for any combination of $\lbrace {\rm
  log}\ t_{\rm Q},\ x_{\rm HeII,0}\rbrace$ between the grid points in
this $2$D parameter space.

\subsection{MCMC}
\label{sec:mcmc}

Having arrived at an expression for the likelihood of each model given
by eqn.~(\ref{eqn:likelihood}) and eqn.~(\ref{eqn:likelihood_full}),
and being able to evaluate the likelihood in any location of the
$\lbrace {\rm log}\ t_{\rm Q},\ x_{\rm HeII,0}\rbrace$ $2$D parameter
space, we can now explore this likelihood with MCMC to determine the
posterior distributions of our parameters. This, in turns, allows us to
estimate the accuracy with which we can measure the mean quasar lifetime
$t_{\rm Q}$ and initial \ion{He}{2} fraction $x_{\rm HeII,0}$ from the
sample of $N=50$ \emph{data} spectra. For these purposes we apply publicly available affine invariant MCMC
ensemble sampling algorithm \emph{emcee} presented and described in \citet{Foreman2013}. 

Following the algorithm described in \S~\ref{sec:like} we perform MCMC
parameter inference on the $N = 50$ \emph{data} spectra drawn from each of 9
different models. The results of this inference
are illustrated in
Figure~\ref{fig:MCMC}, where each panel shows results for each
of the 9 models. The columns
in Figure~\ref{fig:MCMC} show different initial \ion{He}{2}
fractions $x_{\rm HeII,0}^{\rm
  data} = \left[ 0.05, 0.50, 1.00 \right]$, whereas the rows
are for different values of quasar lifetime ${\rm log}\left( t_{\rm Q}^{\rm data}\slash {\rm Myr} \right) = \left[
  0.0, 1.0, 2.0\right]$. The contours illustrate the $95\%$ (blue) and
$68\%$ (red) confidence levels, respectively. Marginalized parameter
distributions for $t_{\rm Q}$ and $x_{\rm HeII,0}$ are also shown by the histograms. We compute the
$16$th, $50$th and $84$th percentiles of these marginalized
distributions, which are quoted as the measurement ($50$th percentile),
and the lower and upper error bars ($16$th and $84$th percentile) in
each panel. There are several notable trends.

First, consider the case where the IGMs initial \ion{He}{2} fraction
is low ($x_{\rm HeII,0}^{\rm data} = 0.05$; left column).  According
to Figure~\ref{fig:TPE_2} there is no significant heating of the gas
due to the quasar turning on, independent of the quasar lifetime
value. Consequently, the power spectrum of these models are not significantly
different and the left column of Figure~\ref{fig:MCMC} illustrates
that the quasar lifetime is essentially unconstrained. Nevertheless,
the lack of a significant thermal proximity effect constrains the
initial \ion{He}{2} fraction to be small, with the typical absolute
errors on $x_{\rm HeII}$ around $\delta_{x_{\rm HeII,0 }} \simeq 0.04$.

Previously we argued that temperature boost is stronger in models with
longer quasar lifetimes and higher values of initial \ion{He}{2}
fraction (see Figure~\ref{fig:TPE_2}). Hence, the thermal broadening
of the \ion{H}{1} Ly$\alpha$ forest follows the same dependence on
$t_{\rm Q}$ and $x_{\rm HeII,0}$, as reflected by the power spectra
in Figures~\ref{fig:PS_1}-\ref{fig:PS_2}. As a result, our MCMC parameter constrains on $t_{\rm Q}^{\rm data}$ and $x_{\rm HeII,0}^{\rm data}$
improve when the thermal proximity effect in the \ion{H}{1} Ly$\alpha$
forest is both larger in size (longer $t_{\rm Q}$), and when the
temperature boost is larger (larger $x_{\rm HeII,0}$). This is illustrated by the middle and right columns of
Figure~\ref{fig:MCMC}, where we show results of MCMC calculations for
two different values of initial \ion{He}{2} fraction, i.e., $x_{\rm HeII,0}^{\rm data} = 0.50$ (middle
column) and $x_{\rm HeII,0}^{\rm data} = 1.00$ (right-hand
column). For instance, because the radial size of the thermal
proximity region is small if the quasar lifetime is only $t_{\rm Q} =
10^6$~yr, the absolute error of the MCMC constraints on the \ion{He}{2}
fraction is $\delta_{x_{\rm HeII,0}} \approx 0.20$ (for $x_{\rm HeII,0}^{\rm data} =
0.50$ and $x_{\rm HeII,0}^{\rm data} = 1.00$). However, as the quasar
lifetime becomes longer (i.e, $t_{\rm Q}^{\rm data} = 10^7$~yr and
$t_{\rm Q}^{\rm data} = 10^8$~yr) the size of the thermal proximity
zone becomes larger, and, hence, the thermal proximity effect is more
prominent. Hence, the mean absolute errors of the MCMC constraints on $x_{\rm HeII}$ shrink down to $\delta_{x_{\rm HeII,0}} \approx 0.10$ if the quasar lifetime is $t_{\rm Q} = 10^7$~yr, and to $\delta_{x_{\rm HeII,0}} \approx 0.05$ in case of $t_{\rm Q} = 10^8$~yr, respectively. Similar trends are apparent for the constraints on quasar lifetime, for which the mean absolute error is $\delta_{{\rm log}t_{\rm Q}} \approx 0.1$~dex when $t_{\rm Q}^{\rm data} \gtrsim 10^7$~yr and $x_{\rm HeII,0}^{\rm data} = 1.00$.

Finally, in \citet{Khrykin2016} we illustrated the degeneracy, which exists between
$t_{\rm Q}$ and the \ion{He}{2} ionizing background $\Gamma_{\rm HeII}^{\rm bkg}$. 
This degeneracy significantly complicates any constraints on $t_{\rm Q}$ or $x_{\rm HeII,0}$ one can obtain from the direct measurements of the \ion{He}{2} proximity zones sizes in far ultraviolet quasar spectra. We argued that this degeneracy can be broken if the value of
$\Gamma_{\rm HeII}^{\rm bkg}$ can be determined from the measurements
of \ion{He}{2} effective optical depth. However, these measurements
become impossible at $z \gtrsim 4$ making it challenging to determine
$t_{\rm Q}$ and $x_{\rm HeII,0}$ from direct observations of
\ion{He}{2} proximity zones. On the other hand, as illustrated in
Section~\ref{sec:T_IGM}, these parameters should not be strongly
degenerate for the thermal proximity effect, because
$t_{\rm Q}$ determines the radial size of the thermal proximity zone and
$x_{\rm HeII,0}$ sets the amplitude of the temperature boost.
Figure~\ref{fig:MCMC} illustrates that this is indeed the case because the contours in each panel of Figure~\ref{fig:MCMC} are aligned with the $x_{\rm HeII}$ and ${\rm log}\ t_{\rm Q}$ directions on the axes.

Given the high sensitivity of our method to the value of the IGM
\ion{He}{2} fraction, our study opens up the exciting possibility of
determining the timing of \ion{He}{2} reionization without direct
UV observations of \ion{He}{2} Ly$\alpha$ absorption at
$z \gtrsim 4$, which are currently impossible. The obvious next
question is how well can we reconstruct the full \ion{He}{2} reionization
history with the thermal proximity effect, which we address in the next section.

\section{Reconstructing the \ion{He}{2} Reionization History}
\label{sec:history}

We apply our thermal proximity effect power spectrum
method and perform MCMC inference on mock datasets extracted from the same set of
hydrodynamical + radiative transfer simulations at $z = \left[ 3.1, 3.5, 3.9, 4.3, 4.6,
  5.0 \right]$ ($1000$ skewers per redshift). Similar to Section~\ref{sec:like}, we construct the same grid of models for each
redshift, covering values of initial \ion{He}{2} fraction $x_{\rm HeII,0} = \left[ 0.05 - 1.00 \right]$ and logarithmically spaced
quasar lifetimes ${\rm log} \left( t_{\rm Q}/{\rm Myr}\right) = \left[ 0.00 - 2.00 \right]$. Figure~\ref{fig:xHeII_Gamma} illustrates the dependence of the average initial \ion{He}{2} fraction at different redshifts on the value of the \ion{He}{2} ionizing background, which sets the initial \ion{He}{2} fraction $x_{\rm HeII,0}$ prior to when the quasar turns on. The value of \ion{H}{1} ionizing background $\Gamma_{\rm HI}^{\rm bkg}$ is adjusted at each redshift in order to match the
mean transmission in the \ion{H}{1} Ly$\alpha$ forest measured by \citet{Becker2013a}. Following the discussion in Section~\ref{sec:like}, we compute the power spectra of all models averaged over $1000$ skewers at each redshift. 

After the grid of average power spectra is computed at all redshifts, the remaining missing ingredient, necessary for the likelihood calculations and MCMC analysis (see Section~\ref{sec:est}), is the sample of modeled spectra representing the \emph{data} at each redshift $z$. For this we adopt the values of $x_{\rm HeII,0}^{\rm data}$ from the fiducial model of \ion{He}{2} reionization from \citet{McQuinn2009} (see model D1 in the bottom panel of their Figure~3), resulting in the following values of \ion{He}{2} fraction in the \emph{data} samples at each redshift
\begin{eqnarray*}
z = 3.1:\ \ x_{\rm HeII,0}^{\rm data} = 0.03;\ z = 4.3:\ \ x_{\rm HeII,0}^{\rm data} = 0.70 \\
z = 3.5:\ \ x_{\rm HeII,0}^{\rm data} = 0.30;\ z = 4.6:\ \ x_{\rm HeII,0}^{\rm data} = 0.80 \\
z = 3.9:\ \ x_{\rm HeII,0}^{\rm data} = 0.50;\ z = 5.0:\ \ x_{\rm HeII,0}^{\rm data} = 0.90
\end{eqnarray*}
We consider three values of quasar lifetime in the \emph{data}
samples, i.e., ${\rm log}\left( t_{\rm Q}^{\rm data}\slash {\rm
  Myr}\right) = \left[ 0.0, 1.0, 2.0 \right]$. Finally, we note that
the number of existing archival high-resolution spectra at $z \simeq
\left[ 4.3, 4.6, 5.0 \right]$ is less than the fiducial number $N =
50$ that we used in our analysis in previous sections. A more
realistic choice is $N = 20$ at these higher redshifts, which we adopt for the \emph{data}
samples at these redshifts, whereas for $z = \left[ 3.1, 3.5, 3.9\right]$
we use the same fiducial number of spectra $N = 50$ as before.

Using eqn.~(\ref{eqn:likelihood})-(\ref{eqn:likelihood_full}) we
perform MCMC parameter inference at each redshift, resulting in
marginalized posterior distributions for $x_{\rm HeII,0}$. Analogous
to Section~\ref{sec:mcmc}, we compute the $16$th, $50$th and $84$th
percentiles of these distributions, and use those to define the
measured value of $x_{\rm HeII,0}^{\rm data}$ ($50$th percentile) and
its uncertainties ($16$th and $84$th percentiles).
Figure~\ref{fig:history} illustrates our reconstructions of the
\ion{He}{2} reionization history, shown by the fraction of completely
ionized helium $x_{\rm HeIII} \equiv 1 - x_{\rm HeII}$. The three
panels correspond to the different values of the quasar lifetime that
were assumed for the \emph{data}, i.e., ${\rm log}\left( t_{\rm
  Q}^{\rm data}\slash {\rm Myr}\right) = 0.0$ (\emph{left}), ${\rm
  log}\left( t_{\rm Q}^{\rm data}\slash {\rm Myr}\right) = 1.0$
(\emph{middle}), and ${\rm log}\left( t_{\rm Q}^{\rm data}\slash {\rm
  Myr}\right) = 2.0$ (\emph{right}), respectively. The red solid
curves in each panel show \citet{McQuinn2009} \ion{He}{2} reionization
history (\emph{red}), whereas the brown solid curve (and grey
uncertainty region) are the default and modified \ion{He}{2}
reionization histories for the semi-analytical reionization model of \citet{Madau2015}, in which the active galactic nuclei dominate the reionization process resulting in an early \ion{He}{2} reionization by the same population of sources that reionized intergalactic hydrogen. The blue dots with error bars are the results of our calculations. It is apparent from Figure~\ref{fig:history} that exploiting the thermal proximity effect in \ion{H}{1} Ly$\alpha$ forest around quasars can be used to fully reconstruct the \ion{He}{2} reionization history, and constrain both its timing and duration.

\section{Thermal Proximity Effect: Wavelet Analysis}
\label{sec:wavelets}

\begin{figure}[t]
\centering
\includegraphics[width=1.0\linewidth]{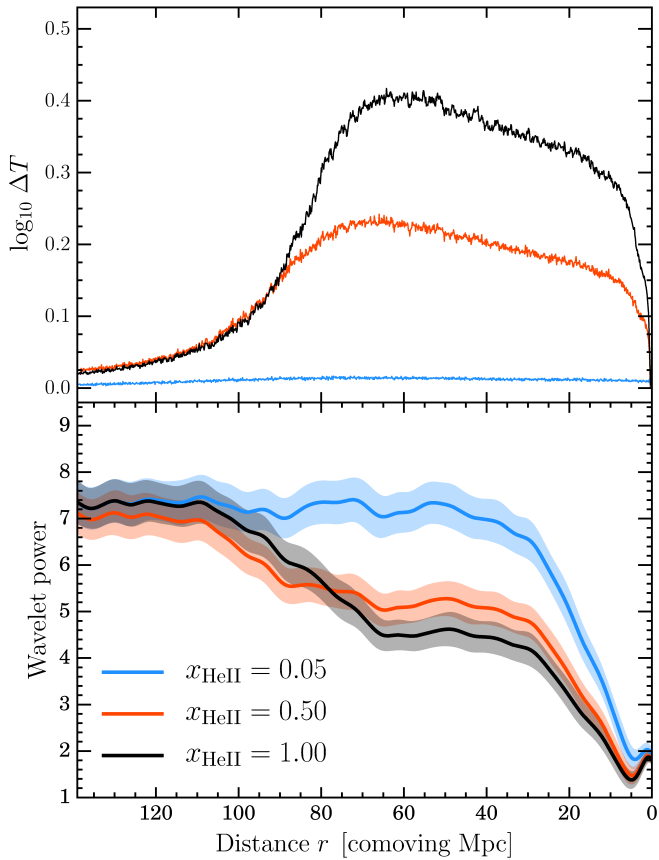}
 \caption{\emph{Upper panel:} Median temperature profiles for three models with $t_{\rm Q} = 10^8$~yr, but different initial \ion{He}{2} fraction: $x_{\rm HeII,0} = 0.05$ (\emph{blue}), $x_{\rm HeII,0} = 0.50$ (\emph{red}), $x_{\rm HeII,0} = 1.00$ (\emph{black}), similar to Figure~\ref{fig:TPE_2}. \emph{Bottom panel:} wavelet coefficients for the same models using $N = 1000$ simulated \ion{H}{1} Ly$\alpha$ spectra. The shaded areas are the $1\sigma$ error bars calculated from $500$ random realizations of $N = 50$ skewers representing the \emph{data} samples drawn from each of the models.}
\label{fig:wavelets}
\end{figure}

Another approach to quantifying the thermal proximity effect is to use
wavelets. Wavelets are filters that are local in both Fourier and configuration space, 
allowing one to select the modes of interest as a function of distance from the quasars. 
There is a rich history of applying wavelets to the Ly$\alpha$ forest to measure the mean
temperature as well as to search for temperature fluctuations
\citep{Theuns2000, Zaldarriaga2002, Lidz2010}.  We can use our
previous power spectrum calculations to design a wavelet to select
modes that contain much of the discriminating information that the
power spectrum analysis is using.  Based on a qualitative analysis of
Figures~\ref{fig:PS_1}-\ref{fig:PS_2}, we chose our wavelet to be a
Gaussian filter centered at $k_0 = 0.13\ {\rm km^{-1}s}$ with a
standard deviation of $\sigma_k = 0.04\ {\rm km^{-1}s}$, roughly the
wavenumbers where the power spectra of the different models for the
thermal proximity effect differ the most.  Note that an offset
Gaussian kernel in Fourier space is a phase times a Gaussian in
configuration space:
\begin{equation}
W_M(x) = \exp \left[ i k_0 x \right] \times \exp \left[ -\frac{x^2\sigma_k^2}{2 }\right].
\end{equation}
This form for the wavelet filter, termed the Morlet wavelet, is convolved with our different models for the flux.  We consider the ``wavelet power'' as a function of distance:
\begin{equation}
  a(r)^2 = \left[ \int W\left( r - r^\prime\right) F\left( r^\prime \right) dr^\prime \right]^2. 
\end{equation}
We then smooth $a \left( r \right)^2$ with a Gaussian kernel with standard deviation $200\,{\rm km\ s^{-1}}$ ($\approx 2.5$~cMpc at $z = 3.9$) to generate a smoothed field, $\langle a \left( r \right)^2 \rangle_{200}$ that is plotted in Figure~\ref{fig:wavelets}. Note that our normalization of $\langle a\left( r \right)^2 \rangle_{200}$ has no physical significance.

Before we discuss the results, we comment on the advantages and
disadvantages of the wavelet approach relative to the power spectrum
analysis used in the rest of the paper. A clear advantage is that the
locality of this filter in configuration space (having width $\sim
1/\sigma_k$ which for our choice is $25\ {\rm km\ s^{-1}}$) has the
nice property that the data do not have to be binned into distance
intervals. This allows one to detect trends with distance
from the quasar, even in the absence of a model. This feature could be
particularly useful well into \ion{He}{2} reionization, when the
heating profile around quasars is likely to be more complicated than we have assumed
\citep{McQuinn2009}. Disadvantages include that (1) it is somewhat
more difficult to estimate posteriors as adjacent wavelet coefficients are correlated, (2) our wavelet, in contrast to the power
spectrum, does not use all discriminating information in 2-point
correlations, and (3), in our experience, it is easier to diagnose
systematics in the power spectrum (as it is easier to diagnose systematics when focusing on a range of wavenumbers).

With these words of caution, we proceed to apply the wavelet to our
models with quasar lifetime fixed at $t_{\rm Q} = 10^8$~yr and
different initial \ion{He}{2} fractions of $x_{\rm HeII,0} = \left[0.05, 0.50, 1.00 \right]$ (the same models as in Figure~\ref{fig:PS_1}). The median of the temperature in these models is shown in the top panel of Figure~\ref{fig:wavelets}. The average wavelet coefficient as a function of distance from the quasar at $z=3.9$ are shown in the bottom panel of Figure~\ref{fig:wavelets}, with each curve calculated from $N = 1000$ mock \ion{H}{1} spectra.  The shaded regions are the $1\sigma$ range from $50$ quasars, calculated using $500$ random realizations. It is apparent, that at large distances all models asymptote to the same power. However, at smaller distances, i.e. $r \lesssim 90\ {\rm cMpc}$ -- the radial extent of the thermal proximity region --, the three models show different wavelet power levels at a level that should be distinguishable with $50$ spectra, and the $x_{\rm HeII,0} = 0.05$ model can be distinguished from the others with far fewer spectra.  Furthermore, the radial trends in the wavelet power trace the temperature and extent of the models' thermal proximity effect.  At smaller radii still, $r \lesssim 20\ {\rm cMpc}$, the effect of the quasar being in an overdense location is apparent in all three cases as a suppression in the wavelength power.\footnote{We note that this effect is likely to be somewhat larger in reality as (1) the quasar locations in our mocks are likely in less biased halos than actual halos, and (2) the scales below which the wavelet power starts to be suppressed is comparable to our simulation box size.} 
  
In conclusion, a wavelet thermal proximity effect analysis is able to distinguish between the models that we consider with a similar number of quasar spectra as that required for a power spectrum analysis.  Since trends in the wavelet coefficient are more easily understood than trends in the power spectrum, any claimed detection of a thermal proximity effect in the power spectrum should be reinforced with a wavelet analysis.  Wavelets further enable a more model-independent search for thermal proximity regions than the analysis advocated in previous sections.

\section{Discussion}
\label{sec:discussion}

In this section we discuss systematics relevant for detecting the thermal proximity effect and placing robust constraints on \ion{He}{2} fraction and quasar lifetime. In addition, we discuss several assumptions that went into our previous estimates and how they affect our conclusions.

\subsection{The Uniformity of the \ion{He}{2} Ionizing Background}

Throughout this work we have assumed that the \ion{He}{2} ionizing
background is constant in space and time, and added it in each pixel
along simulated skewers.  Here we note that, except for possibly when
$x_{\rm HeII} \ll 1$, a pervasive and uniform background flux is
likely not a good approximation. Our methodology of including a
uniform background photoionization rate is a convenience to maintain
the desired $x_{\rm HeII}$ at the mean density.  Since the \ion{He}{3}
$\rightarrow$ \ion{He}{2} recombination time is comparable to the age
of the Universe, the effect of this background on our results is
minimal and our results would not change if we turned it off (as
motivated in \S~\ref{sec:model}, we also do not include the
photoheating from this background).  We also note that when $x_{\rm
  HeII}$ is appreciably different from unity, we expect each quasar to
be turning on in a swiss cheese of relic \ion{He}{3} regions. The
approximation of a uniform $x_{\rm HeII}$ is meant to describe the
average effect.  Since our analysis ultimately stacks measurements
from many quasars and since we are concerned with reionization and
heating -- which are linear in $\Delta x_{\rm HeII}$ (see
eqn.~\ref{eqn:temp_evol_final}) -- we expect the thermal proximity profile in a stack
of real sightlines to be reasonably approximated by the stack of
quasars going off in a medium with uniform ionizing background and
$x_{\rm HeII}$.\footnote{Since quasars go off in biased locations in
  the Universe, this means that the typical $x_{\rm HeII}$ will be
  smaller nearer to quasars. However, since quasar bubbles are so
  large, simulations of \ion{He}{2} reionization suggest that such correlations are weak
  and the fluctuations in $x_{\rm HeII}$ are reasonably approximated by
  randomly throwing down \ion{He}{3} bubbles \citep{McQuinn2009, Compostella2013} .}

\subsection{Initial Thermal State of the IGM}

As discussed at the beginning of \S~\ref{sec:like}, our calculations assume perfect knowledge of the thermal state of the
IGM before the quasar turns on, and that the hydrodynamical
simulations we compare to have been calibrated to reproduce this
thermal state of the ambient IGM using data far from quasars. However,
note that the thermal state of the ambient IGM is set by the
\ion{He}{2} reionization history. For example, a quasar turning on in
an IGM which has an average $x_{\rm HeII,0} = 0.05$ implies that
\ion{He}{2} was already reionized by an earlier generation of quasars
(or other sources). In this case, the IGM around the quasar should
have already been photoheated to higher temperature by the earlier
reionization of \ion{He}{2}. Therefore for a model with $x_{\rm HeII,0} = 0.05$,
quasars would turn on in an IGM which is actually initially hotter
than a $x_{\rm HeII,0} = 1.00$ model, for which there was not yet
any \ion{He}{2} reionization heating. This $x_{\rm HeII,0}$ dependent
initial thermal state has not been modeled in our calculations.

The sensitivity of the power spectrum analysis presented here depends
on the amount of heat $\Delta T$ that was injected into the IGM by
quasar radiation (see discussion in \S~\ref{sec:powspec}) relative to the initial
temperature of the IGM. Therefore, our analysis is likely sensitive to the initial 
IGM thermal state. If the temperature of the ambient IGM is much higher (lower) than what we have
assumed in our models, the precision of our MCMC constraints in
\S~\ref{sec:mcmc} could be over-estimated (under-estimated). Throughout this work we used the outputs of hydrodynamical simulations with the value of ambient IGM temperature $T_0 \simeq 10^4$~K, which is probably too low in case when initial \ion{He}{2} fraction is $x_{\rm HeII,0} = 0.05$ as outlined previously.

\begin{figure*}[t]
\centering
\includegraphics[width=0.9\linewidth]{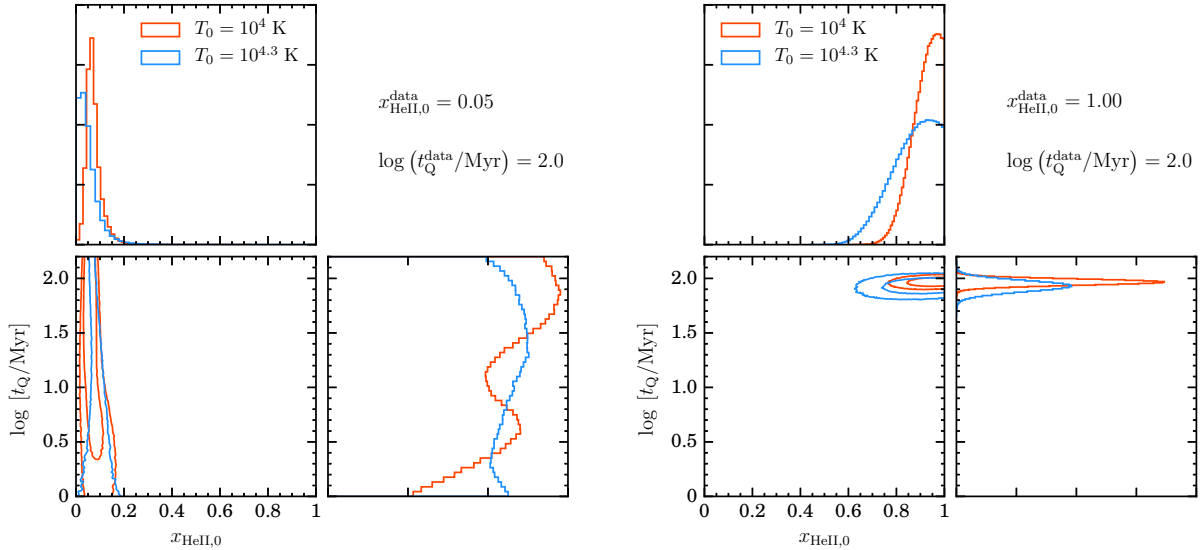}
 \caption{Constraints on $x_{\rm HeII,0}$ and $t_{\rm Q}$ from the MCMC analysis in case of increased initial IGM temperature for $x_{\rm HeII,0}^{\rm data} = 0.05$ (\emph{left}) and $x_{\rm HeII,0}^{\rm data} = 1.00$ (\emph{right}). The bottom left subpanels illustrate the $95\%$ and $68\%$ confidence intervals on $x_{\rm HeII,0}$ and $t_{\rm Q}$ for two cases: (1) analogous to Figure~\ref{fig:MCMC} when $T_0 \simeq 10^4$~K (\emph{red}), and (2) increased initial IGM temperature $T_0 \simeq 10^{4.3}$~K (\emph{blue}). The top and bottom right subpanels show the marginalized parameter distributions.}
\label{fig:mcmc_T0}
 \end{figure*} 

For this reason and in order to verify the sensitivity of our analysis to the level of
initial IGM temperature, we increase $T$ in all pixels along $1000$
skewers drawn from hydrodynamical simulations at $z = 3.9$ by
$\Delta T = 7000^{\circ}$K, and then run our radiative transfer calculations and construct the same grid of models as in
\S~\ref{sec:like}. Adding this $\Delta T$ mimics the heating produced by \ion{He}{2}
reionization, naturally flattening the temperature-density relation as
expected due to the roughly density independent injection of heat (see \S~\ref{sec:T-rho}).
It also leaves temperatures effectively unchanged in hot shock-heated regions. Similar to the discussion in \S~\ref{sec:like}, we
estimate the likelihood of two \emph{data} samples of $N = 50$ skewers
drawn from the models with the same value of quasar lifetime $t_{\rm
  Q}^{\rm data} = 10^8$~yr, but different initial \ion{He}{2}
fractions, i.e., $x_{\rm HeII,0}^{\rm data} = 0.05$ and $x_{\rm
  HeII,0}^{\rm data}=1.0$. We explore these likelihoods with our MCMC
analysis (see \ref{sec:mcmc}).

The results of this test are shown in Figure~\ref{fig:mcmc_T0}, where we compare the results of MCMC analysis in case of increased initial temperature of the IGM to those we obtained previously in \S~\ref{sec:mcmc}. The left side panels show the results in case $x_{\rm HeII,0}^{\rm data} = 0.05$, whereas the right side panels are for $x_{\rm HeII,0}^{\rm data} = 1.0$. It is apparent from Figure~\ref{fig:mcmc_T0} that, similar to the bottom left panel in
Figure~\ref{fig:MCMC}, when $x_{\rm HeII,0}^{\rm data} = 0.05$ there
is no sensitivity to quasar lifetime due to lack of thermal proximity
effect, but the constraints we obtain on initial \ion{He}{2} fraction
given increased initial temperature of the IGM ($x_{\rm HeII,0} =
0.04^{+0.04}_{-0.02}$) are comparable to those in
\S~\ref{sec:mcmc}. On the other hand, Figure~\ref{fig:mcmc_T0}
illustrates that in case $x_{\rm HeII,0}^{\rm data} = 1.0$ the
constraining power of our analysis is decreased by $\simeq 50\%$ when
compared to results in \S~\ref{sec:mcmc}. This is due to the fact that
the heat injection $\Delta T$ (see eqn.~\ref{eqn:temp_evol_final}) is
smaller, hence the thermal proximity effect in the \ion{H}{1}
Ly$\alpha$ forest is weaker. However, keep in mind that we have actually modeled a rather extremely high temperature for
$x_{\rm HeII,0}^{\rm data} = 1.00$ case (as our initial temperature of $T_0 \simeq 10^{4.3}$~K is motivated by expectations for when most of the \ion{He}{2} had been reionized).

Lastly, our analysis largely ignores the complicated radiative transfer
and heating that is expected during the last half of \ion{He}{2}
reionization as the ionizing regions largely overlap and hard photons
heat locations far from quasars \citep{McQuinn2009, Compostella2013}. The trends with radius should
be weaker than at the earlier times studied here, when proximate
heating dominates the radial trend. Modeling the morphology of this heating
likely requires full $3$D calculations, and non-parametric wavelet methods
are likely more suited to this more model dependent case.

\subsection{Continuum Placement and the Mean Flux}
\label{sec:cont_mean_f}

Another systematic that could affect the thermal proximity effect
inferences is the placement of the quasar continuum.  Throughout this
work we have assumed perfect knowledge of the quasar continuum, but
misplacement of the quasar continuum could lead to systematic errors
in the power spectrum measurements which are at the heart of our
method to measure the thermal proximity effect.  Although continuum
fitting uncertainties are relatively small ($\sim$ few per cent) at $z
\simeq 2-3$ using high-resolution data, they can reach $\simeq 10 -
20\%$ at $z \gtrsim 4$ \citep{FG2008}. Furthermore, the thermal
proximity effect falls on the tail of the quasar's broad Ly$\alpha$
line, thus the continuum errors could be significantly larger at this
location in the spectrum.

To explore how the continuum placement affects power spectrum
estimates, let us define a different flux contrast field $\delta
F_{\rm C}$ as follows
\begin{equation}\label{eqn:flux_contr2}
\delta F_{\rm C} \left( r \right) = \frac{F(r) - \langle F \rangle_{\rm local}} {\langle F \rangle_{\rm local}} = \frac{f_{\rm obs}(r)\slash C - \langle f_{\rm obs}\slash C \rangle_{\rm local}} {\langle f_{\rm obs}\slash C \rangle_{\rm local}}
\end{equation}
where $C$ is the continuum flux, $F$ is the transmitted flux, and
$\langle F \rangle_{\rm local} = \langle f_{\rm obs} \slash C
\rangle_{\rm local}$ is the mean flux in the $10$~cMpc radial bin
under consideration. Note that $\langle F \rangle_{\rm local}$ is
estimated not from an ensemble of quasars, but rather from each
individual quasar spectrum. Whereas our original definition of flux
contrast $\delta F$ (eqn.~\ref{eqn:flux_contr}) implicitly assumed
perfect knowledge of the continuum and the mean flux, this new $\delta
F_{\rm C}$ suffers from additional noise, because $\langle F
\rangle_{\rm local}$ is a noisy estimate of the true mean flux
$\langle F\rangle$ due to fluctuations on $\sim 10$~cMpc scales. This method of local continuum fitting is analogous to the
  ``trend-removal'' approach adopted in previous Ly$\alpha$ forest studies \citep{Hui2001, Croft2002, Lidz2006, Lidz2010}. 
However, if the continuum $C$ is roughly constant over this $10$~cMpc
$r$-bin, then according to the definition of $\delta F_{\rm C}$ in
eqn.~(\ref{eqn:flux_contr2}), $C$ cancels out and the power
spectrum is insensitive to the continuum level.  Even if $C$ is not
exactly constant, it is very likely that the continuum does not contribute much
power on scales smaller than $10$~cMpc, and so our new $\delta F_{\rm C}$ will
be insensitive to continuum errors. We now redo our analysis to see if
the reduced information in $\delta F_{\rm C}$
(eqn.~\ref{eqn:flux_contr2}) relative to $\delta F$
(eqn.~\ref{eqn:flux_contr}) affects our results.

\begin{figure}[b]
\centering
\includegraphics[width=0.8\linewidth]{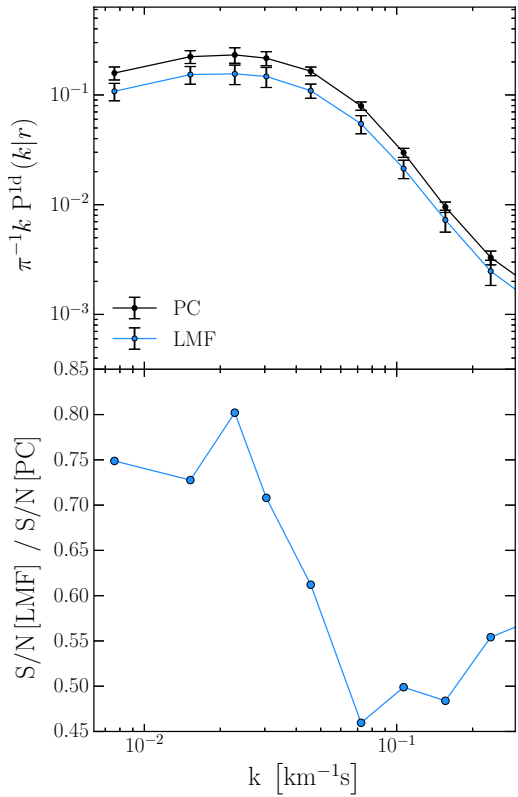}
 \caption{Comparison between the power spectra of the \ion{H}{1} Ly$\alpha$ forest at $r = 43-53$~cMpc from the quasar, calculated in two methods. In the upper panel the solid \emph{black} line shows the case when the perfect knowledge of the quasar continuum (``PC") is assumed (similar to black curve in the middle panel of Figure~\ref{fig:PS_1}), while the \emph{blue} curve illustrates the results of the local mean flux (``LMF") analysis (see discussion in the text for details). In both cases the power spectra are computed from $1000$ skewers drawn from the model with $t_{\rm Q} = 10^8$~yr and $x_{\rm HeII,0} = 1.00$. The errorbars are computed from 500 random realizations of \emph{data} power spectrum drawn from original 1000 skewers of the each model. The bottom panel shows the ratio of signal-to-noise ratios in two cases.}
\label{fig:mftest}
 \end{figure} 

Using the new flux contrast $\delta F_{\rm C}$, we perform our calculations from \S~\ref{sec:est} for the case
of $N = 50$ skewers, with $t_{\rm Q}^{\rm data} = 10^8$~yr, and $x_{\rm HeII,0}^{\rm data} = 1.0$. We compute the average power spectrum
$P_{\rm data}\left(k|r\right)\equiv \langle |\delta \tilde{F}_{\rm C}(k) |^2 \rangle$. We then calculate the likelihood of this \emph{data} sample in each radial bin we consider, and perform parameter inference with MCMC. The results of this modified analysis, to which we refer as ``LMF" (Local Mean Flux), are shown in
Figure~\ref{fig:mftest}, where we present power spectra of the \ion{H}{1} Ly$\alpha$ forest in the $r = 43 - 53$~cMpc radial bin of the
thermal proximity region (see middle panel of Figures~\ref{fig:PS_1}-\ref{fig:PS_2}).

\begin{figure}[t]
\centering
\includegraphics[width=1.0\linewidth]{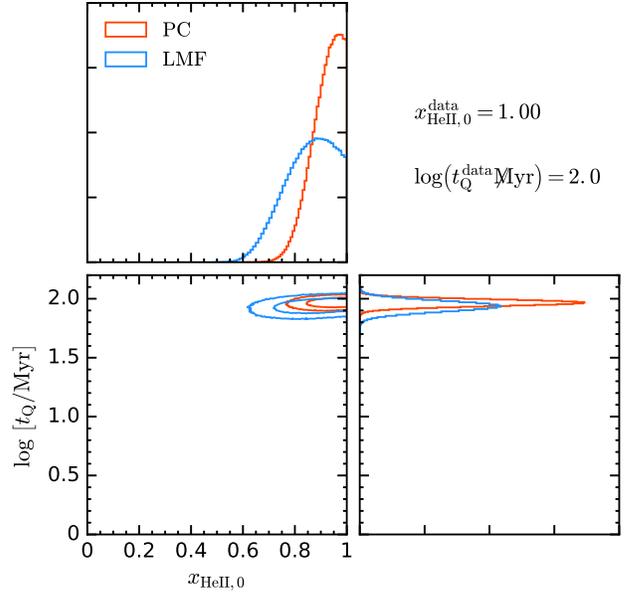}
 \caption{Comparison between the constraints on $x_{\rm HeII,0}$ and $t_{\rm Q}$ obtained from the MCMC analysis in two methods. The bottom left panel shows the $95\%$ and $68\%$ confidence intervals on $x_{\rm HeII,0}$ and $t_{\rm Q}$ in case when (1) similar to Figure~\ref{fig:MCMC}, the perfect knowledge of the quasar continuum is assumed (``PC"), which is shown in \emph{red}, and (2) when the local mean flux (``LMF") is used in models instead, which is shown in \emph{blue}. The top and bottom right panels show the marginalized parameter distributions for $x_{\rm HeII,0}$ and ${\rm log}\ t_{\rm Q}$, respectively. The solid \emph{red} histograms are for the PC case, while the solid \emph{blue} ones illustrate the LMF approximation.}
\label{fig:mftest_mcmc}
 \end{figure} 

The black curve in the upper panel of Figure~\ref{fig:mftest}
shows the power spectrum of the $t_{\rm Q} = 10^8$~yr, $x_{\rm
  HeII,0} = 1.0$ model (averaging $1000$ skewers), calculated
assuming perfect knowledge of the quasar continuum (the same as in
\S~\ref{sec:powspec}, which we refer to as ``PC" - Perfect Continuum).  The blue curve shows the power spectrum
of the same model calculated using the LMF (eqn.~\ref{eqn:flux_contr2}). The errorbars are computed from $500$ random realizations of \emph{data} power spectrum drawn from original $1000$ skewers of each model. The bottom panel shows the ratio of the signal-to-noise ratio (S/N) for the two methods defined as $\Delta = [ P_{\rm model} \slash \sigma ]_{\rm LMF} \slash [ P_{\rm model} \slash \sigma ]_{\rm PC}$. The power spectrum and the error bars
on the \emph{data} power spectrum are reduced when we calculate the mean flux locally.  This is apparent from the bottom panel of Figure~\ref{fig:mftest},
where one can see that at $k \gtrsim 2 \times 10^{-2}\ {\rm km^{-1}s}$ the ${\rm S\slash N}$ ratio has dropped by $\approx 50\%$ for the LMF case relative to the previous results of \S~\ref{sec:powspec}.
 
Figure~\ref{fig:mftest_mcmc} shows the results of MCMC calculations for the same \emph{data} sample for two methods, i.e., the \emph{red} contours illustrate the results for the global mean flux used in the power spectrum calculations, whereas the \emph{blue} contours are for the LMF case. It is apparent that our analysis of the thermal proximity effect has lost about half of its constraining power, but nevertheless still allows robust constraints on both quasar lifetime $t_{\rm Q}$ and initial \ion{He}{2} fraction $x_{\rm HeII,0}$. The reason why the constraints have weakened by this amount is that the LMF method has made us insensitive to the thermal proximity effect's impact on the average transmission profile (remember $x_{\rm HI} \propto T^{-0.7}$ and so the mean absorption in the forest is modulated by the thermal proximity effect), and because $\delta F_{\rm C}$ suffers from additional noise due to fluctuations of $\langle F \rangle_{\rm local}$ about the true mean flux.

To conclude, we have presented two distinct cases. First, we assumed
perfect knowledge of the quasar continuum, which gives the best
possible constraints. Second, we analyzed the \emph{worst} case
scenario when there is no knowledge of the quasar continuum in the
thermal proximity region. In this case, despite the drop in overall
constraining power of our method, we were still able to constrain the
parameters of interest. Note however that in principle, one could
jointly model both the power spectrum and the mean flux profile in the
thermal proximity zone, effectively mitigating the extra noise present
in the LMF case.

\subsection{Degeneracy with Photon Production Rate $Q_{\rm 4Ry}$}
\label{sec:lum_cav}

In \S~\ref{sec:temp_prof} we have illustrated how the radial size of the thermal proximity region depends on the quasar lifetime $t_{\rm Q}$. In practice, the size of this region is correlated with the distance $r_{\rm IF}$ to which the quasar ionization front traveled in time $t = t_{\rm Q}$. Given that the recombination time is long compared to the quasar lifetimes we consider in this work ($t_{\rm Q} \ll t_{\rm rec} \simeq 10^9$~yr), the location of the quasar ionization front is  approximately proportional to the product of quasar lifetime and quasar photon production rate, i.e., $r_{\rm IF} \propto \left( t_{\rm Q} \times Q_{\rm 4Ry} \right)^{1/3}$. Hence, in practice, the radial size of the thermal proximity region must depend on a degenerate combination of $t_{\rm Q}$ and $Q_{\rm 4Ry}$, or the total number of ionizing photons emitted. 

However, in this work we have ignored this degeneracy by fixing $Q_{\rm 4Ry}$ to a single value and illustrated the effect of quasar lifetime only. While this might seem to be an unnecessary simplification, we note that, in reality, the degeneracy between $t_{\rm Q}$ and $Q_{\rm 4Ry}$ can be broken. This is because the average quasar SED directly constraints the quasar apparent magnitude $m_{912}$ at the Lyman limit $\lambda = 912 {\rm \AA}$, which, given the knowledge of the quasar SED slope $\alpha_{\rm 1Ry \rightarrow 4Ry}$ between $1$~Ry and $4$~Ry \citep{Telfer2002, Shull2012, Lusso2015}, can be used to estimate the quasar photon production rate at $4$~Ry $Q_{\rm 4Ry}$. The most uncertain thing is the slope of quasar SED above $4$~Ry $\alpha_{\rm 4Ry \rightarrow \infty}$, which can change $Q_{\rm 4Ry}$ by $\approx 25-45\%$ \citep{Khrykin2016}. However, the $z \simeq 3$ \ion{He}{2} Ly$\alpha$ proximity effect can be used to calibrate $\alpha_{\rm 4Ry \rightarrow \infty}$ using stacked proximity zone profiles \citep{Khrykin2016}. Note also that although quasars have different luminosities, we are sensitive to the average quantities only in our analysis of the thermal proximity effect. Thus, we argue that constraints on the average quasar luminosity that can be calibrated with the observations, will allow one to determine the average quasar lifetime using the thermal proximity effect.

\subsection{The Dependence on the Spectral Slope $\alpha_{\nu}$}
\label{sec:alpha_cav}

\begin{figure}[!t]
\centering
 \includegraphics[width=0.8\linewidth]{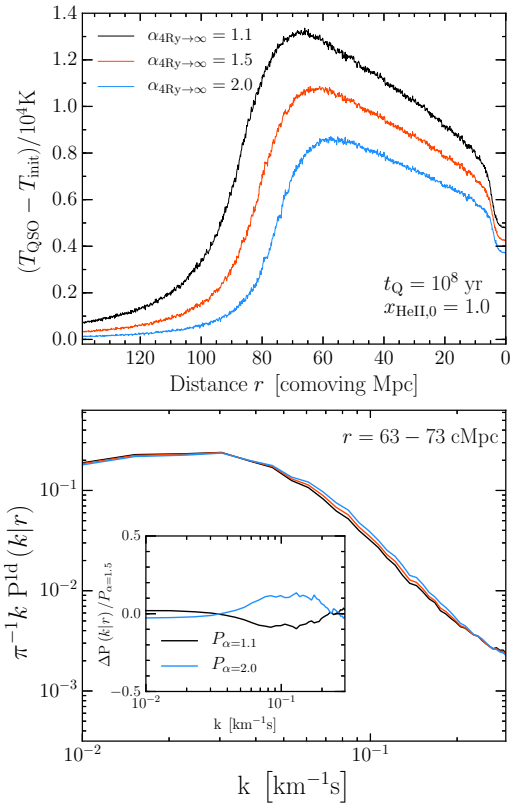}
 \caption{Thermal evolution of the intergalactic medium around the quasar in different radiative transfer simulations. \emph{ The upper panel} shows the evolution of the median IGM temperature profiles (similar to the bottom panels of Figure~\ref{fig:TPE_2}) as a function of the slope of the quasar SED at $\nu \geq \nu_{\rm 4Ry}$ $\alpha_{\rm 4Ry \rightarrow \infty}$, whilst \emph{the bottom panel} illustrates the resulting average power spectra of the same models in $r = 63-73$~cMpc radial bin. The embedded panel shows the percent change in the power spectra compared to the power spectrum of our fiducial model with $\alpha_{\rm 4Ry \rightarrow \infty} = 1.5$. Each curve is a stack of a $1000$ skewers. The quasar lifetime and the initial \ion{He}{2} fraction in each model are $t_{\rm Q} = 10^8$~yr and $x_{\rm HeII,0} = 1.0$, respectively. }
\label{fig:temp_alpha}
\end{figure}

Throughout the paper we have assumed a constant slope of the quasar SED $\alpha_{\nu}$ at frequencies blueward of $1$~Ry, with $\alpha_{\nu} = 1.5$. However, the spectral slope regulates the number of high-energy photons with long mean free path, therefore any change in $\alpha_{\nu}$ can affect the IGM temperature and \ion{H}{1} transmission at large distances from the quasar (see discussion in \S~\ref{sec:T_IGM}). As we discussed in \S~\ref{sec:lum_cav}, the photon production rate $Q_{\rm 4Ry}$ can be constrained from observations given our good knowledge of the spectral slope between $1$ and $4$~Ry. However, the slope beyond $4$~Ry is currently not well defined. In order to investigate how variations in the slope of the quasar SED above $4$~Ry affect our results, we have run an additional set of radiative transfer simulations, and in what follows we describe our findings.

Analogous to the discussion in \citet{Khrykin2016}, we fix the quasar specific photon production rate $N_{\rm 4Ry}$ (see eqn.~\ref{eqn:lum}), which is calculated from the observable specific photon production rate $N_{\rm 1Ry}$ at $1$~Ry extrapolated as a power law to $4$~Ry assuming a spectral slope $\alpha_{\rm 1Ry \rightarrow 4Ry} = 1.5$, consistent with recent constraints from quasar SED \citep{Telfer2002, Shull2012, Lusso2015}\footnote{This spectral slope $\alpha_{\rm 1Ry \rightarrow 4Ry}$ can also be calibrated from the \ion{H}{1} Ly$\alpha$ line-of-sight proximity effect.}. We then change the SED slope from $4$~Ry to infinity within the range $\alpha_{\rm 4Ry \rightarrow \infty} = 1.1 - 2.0$\footnote{This range is motivated by the current constraints on the quasar SED slope from \citet{Lusso2015}, who found $\alpha_{\nu} = 1.70 \pm 0.61$ at $\lambda \leq 912 {\rm \AA}$.}, and run our radiative transfer algorithm.

The upper panel of Figure~\ref{fig:temp_alpha} illustrates the effect of varying the slope of quasar SED $\alpha_{\rm 4Ry \rightarrow \infty}$ on the median IGM  temperature profiles. The quasar lifetime and initial \ion{He}{2} fraction are set to $t_{\rm Q} = 10^8$~yr and $x_{\rm HeII,0} = 1.00$ in all models. As expected (see discussion in \S~\ref{sec:T_IGM}), smaller (larger) values of $\alpha_{\rm 4Ry \rightarrow \infty}$ increase (decrease) the number of hard photons produced by the quasar, and hence change the amount of energy injected into the IGM, dramatically altering the amplitude of the thermal proximity effect. Namely, the maximum temperature boost is a factor of $2$ higher (lower) in case when $\alpha_{\rm 4Ry \rightarrow \infty} = 1.1$ ($\alpha_{\rm 4Ry \rightarrow \infty} = 2.0$), compared to the fiducial model with $\alpha_{\rm 4Ry \rightarrow \infty} = 1.5$. It is also apparent that the peak of the temperature profile shifts toward larger distances from the quasar when the spectral slope becomes harder (i.e., $\alpha_{\rm 4Ry \rightarrow \infty} = 1.1$). Since the mean free path of ionizing photons scales as $\lambda_{\rm mfp} \propto \nu^{3}$, the high-energy photons can travel further into the IGM, depositing their energy at larger distances, and because there are more such photons for a harder SED shape ($\alpha_{\rm 4Ry \rightarrow \infty} = 1.1$), the peak shifts.

On the other hand, the bottom panel of Figure~\ref{fig:temp_alpha} illustrates the resulting average line-of-sight \ion{H}{1} power spectra for the same models. It is apparent that the variations in $\alpha_{\rm 4Ry \rightarrow \infty}$ result in only modest differences in the power spectrum ($\Delta P/P_{\alpha = 1.5} \simeq 10\%$). These are substantially smaller than the uncertainties in the continuum placement discussed in \S~\ref{sec:cont_mean_f}. Hence, we argue that variations in $\alpha_{\rm 4Ry \rightarrow \infty}$ should not affect our analysis of the thermal proximity effect significantly.

Finally, we remind the reader that the slope of quasar SED above $4$~Ry can be calibrated with the lower-$z$ \ion{He}{2} Ly$\alpha$ proximity zone profiles \citep{Khrykin2016}. Alternatively, it is also possible to constrain the spectral index $\alpha_{\rm 4Ry \rightarrow \infty}$ by modeling the \ion{He}{2} ionizing background at $2.5 \leq z \leq 3.5$ and comparing it to $\Gamma_{\rm HeII}^{\rm bkg}$ estimated from measurements of \ion{He}{2} effective optical depth at the same redshifts \citep{Khaire2017}.

\section{Conclusions}
\label{sec:conclusions}

We combined cosmological hydrodynamical simulations with $1$D
radiative transfer calculations to investigate the line-of-sight thermal proximity
effect and its detectability in \ion{H}{1} Ly$\alpha$ forest
absorption spectra. The hard radiation emitted by quasars photoionizes
\ion{He}{2} in their environment, and the resulting photoelectric
heating boosts the temperature of the surrounding IGM. We showed how
the radial temperature profile around quasars depends on the quasar
lifetime, $t_{\rm Q}$, and the average \ion{He}{2} fraction, $x_{\rm HeII,0}$, 
in the IGM before the quasar turned on.  The
main conclusions of our work are:

\begin{enumerate}

\item The amplitude of the thermal proximity effect depends strongly on the average
  amount of singly ionized helium, which prevailed in the IGM prior to the quasar
  activity, $x_{\rm HeII,0}$, whereas the size of the thermal proximity zone
  depends on the average quasar lifetime, $t_{\rm Q}$. 

\item We presented a new method to detect this temperature boost, and
  thus constrain $x_{\rm HeII,0}$ and $t_{\rm Q}$, by measuring the
  \ion{H}{1} power spectrum of an ensemble of quasar spectra as a
  function of distance from the quasar (in several $\sim 10$~cMpc
  bins), and comparing the results to control regions far away from
  the quasar. We also discussed another method based on the wavelet analysis, which enables a non-parametric study of the thermal proximity effect.

\item Combining our power spectrum method with a Bayesian MCMC
  formalism, we showed that a mock dataset of $50$ quasars at $z \sim
  4$ can be used to measure the mean quasar lifetime $t_{\rm Q}$ to a
  precision of $\sim 0.1$~dex, and the initial fraction of singly
  ionized helium $x_{\rm HeII,0}$ to an (absolute) precision of up to
  $\sim 0.05$.

\item Our calculations show that existing \ion{H}{1} Ly$\alpha$ forest datasets can use line-of-sight thermal proximity effect to reconstruct the full
  \ion{He}{2} reionization history over the
  redshift range $3.1 \leq z \leq 5.0$, constraining the timing and
  duration of \ion{He}{2} reionization.

\item We discussed several sources of uncertainties that affect
  the constraining power of our analysis, including the initial
  thermal state of the IGM and uncertainties in the
  continuum spectra of quasars. We found that even under the most extreme assumptions, the
  constraining power of our method decreases by at most a factor of $\approx 2$. 

\end{enumerate}

Reconstruction of the \ion{He}{2} reionization history over $3 \lesssim z \lesssim 5$ with our method would in turn
provide a global census of hard ($> 4$ Rydberg) ionizing photons in
the high-$z$ Universe. Besides being fundamental for understanding
\ion{He}{2} reionization, this census would have important
implications for the thermal state of the IGM, the luminosity density
of quasars and AGN (or other sources of hard photons), possibly even \ion{H}{1} reionization  \citep{Madau2015, DAloisio2016}. 

\section{Acknowledgements}

We would like to thank members of the ENIGMA\footnote{http://enigma.physics.ucsb.edu/} group at the
Max-Planck-Institut f{\"u}r Astronomie (MPIA) for useful discussions and comments on the paper. We are grateful to the anonymous referee for comments and suggestions, which greatly improved the text. I.S.K. thanks Martin Haehnelt for useful discussion about the mean flux issues and Michael Walther for his assistance. I.S.K. acknowledges support from the grant of the Ministry of Education and Science of the Russian Federation \textnumero $3.858.2017$. M.M. acknowledges support from NSF through grant AST 1514734 and NASA through grant HSTAR-13903.00.

\bibliography{ref}

\begin{thebibliography}{}
\expandafter\ifx\csname natexlab\endcsname\relax\def\natexlab#1{#1}\fi

\bibitem[{{Abel} \& {Haehnelt}(1999)}]{Abel1999}
{Abel}, T., \& {Haehnelt}, M.~G. 1999, \apjl, 520, L13

\bibitem[{{Anderson} {et~al.}(1999){Anderson}, {Hogan}, {Williams}, \&
  {Carswell}}]{Anderson1999}
{Anderson}, S.~F., {Hogan}, C.~J., {Williams}, B.~F., \& {Carswell}, R.~F.
  1999, \aj, 117, 56

\bibitem[{{Araya} \& {Padilla}(2014)}]{Araya2014}
{Araya}, I.~J., \& {Padilla}, N.~D. 2014, \mnras, 445, 850

\bibitem[{{Bajtlik} {et~al.}(1988){Bajtlik}, {Duncan}, \&
  {Ostriker}}]{Bajtlik1988}
{Bajtlik}, S., {Duncan}, R.~C., \& {Ostriker}, J.~P. 1988, \apj, 327, 570

\bibitem[{{Becker} {et~al.}(2011){Becker}, {Bolton}, {Haehnelt}, \&
  {Sargent}}]{Becker2011}
{Becker}, G.~D., {Bolton}, J.~S., {Haehnelt}, M.~G., \& {Sargent}, W.~L.~W.
  2011, \mnras, 410, 1096

\bibitem[{{Becker} {et~al.}(2013a){Becker}, {Hewett}, {Worseck}, \&
  {Prochaska}}]{Becker2013a}
{Becker}, G.~D., {Hewett}, P.~C., {Worseck}, G., \& {Prochaska}, J.~X. 2013a,
  \mnras, 430, 2067

\bibitem[{{Bolton} {et~al.}(2012){Bolton}, {Becker}, {Raskutti}, {Wyithe},
  {Haehnelt}, \& {Sargent}}]{Bolton2012}
{Bolton}, J.~S., {Becker}, G.~D., {Raskutti}, S., {et~al.} 2012, \mnras, 419,
  2880

\bibitem[{{Bolton} {et~al.}(2010){Bolton}, {Becker}, {Wyithe}, {Haehnelt}, \&
  {Sargent}}]{Bolton2010}
{Bolton}, J.~S., {Becker}, G.~D., {Wyithe}, J.~S.~B., {Haehnelt}, M.~G., \&
  {Sargent}, W.~L.~W. 2010, \mnras, 406, 612

\bibitem[{{Bolton} {et~al.}(2009){Bolton}, {Oh}, \& {Furlanetto}}]{Bolton2009}
{Bolton}, J.~S., {Oh}, S.~P., \& {Furlanetto}, S.~R. 2009, \mnras, 395, 736

\bibitem[{{Borisova} {et~al.}(2015){Borisova}, {Lilly}, {Cantalupo},
  {Prochaska}, {Rakic}, \& {Worseck}}]{Borisova2015}
{Borisova}, E., {Lilly}, S.~J., {Cantalupo}, S., {et~al.} 2015, ArXiv e-prints,
  arXiv:1510.00029

\bibitem[{{Compostella} {et~al.}(2013){Compostella}, {Cantalupo}, \&
  {Porciani}}]{Compostella2013}
{Compostella}, M., {Cantalupo}, S., \& {Porciani}, C. 2013, \mnras, 435, 3169

\bibitem[{{Compostella} {et~al.}(2014){Compostella}, {Cantalupo}, \&
  {Porciani}}]{Compostella2014}
---. 2014, \mnras, 445, 4186

\bibitem[{{Croft} {et~al.}(2002){Croft}, {Weinberg}, {Bolte}, {Burles},
  {Hernquist}, {Katz}, {Kirkman}, \& {Tytler}}]{Croft2002}
{Croft}, R.~A.~C., {Weinberg}, D.~H., {Bolte}, M., {et~al.} 2002, \apj, 581, 20

\bibitem[{{D'Aloisio} {et~al.}(2016){D'Aloisio}, {Upton Sanderbeck}, {McQuinn},
  {Trac}, \& {Shapiro}}]{DAloisio2016}
{D'Aloisio}, A., {Upton Sanderbeck}, P.~R., {McQuinn}, M., {Trac}, H., \&
  {Shapiro}, P.~R. 2016, ArXiv e-prints, arXiv:1607.06467

\bibitem[{{Davies} {et~al.}(2016){Davies}, {Furlanetto}, \&
  {McQuinn}}]{Davies2016}
{Davies}, F.~B., {Furlanetto}, S.~R., \& {McQuinn}, M. 2016, \mnras, 457, 3006

\bibitem[{{Fan} {et~al.}(2006){Fan}, {Strauss}, {Becker}, {White}, {Gunn},
  {Knapp}, {Richards}, {Schneider}, {Brinkmann}, \& {Fukugita}}]{Fan2006}
{Fan}, X., {Strauss}, M.~A., {Becker}, R.~H., {et~al.} 2006, \aj, 132, 117

\bibitem[{{Faucher-Gigu{\`e}re} {et~al.}(2008){Faucher-Gigu{\`e}re},
  {Prochaska}, {Lidz}, {Hernquist}, \& {Zaldarriaga}}]{FG2008}
{Faucher-Gigu{\`e}re}, C.-A., {Prochaska}, J.~X., {Lidz}, A., {Hernquist}, L.,
  \& {Zaldarriaga}, M. 2008, \apj, 681, 831

\bibitem[{{Finkelstein} {et~al.}(2012){Finkelstein}, {Papovich}, {Ryan},
  {Pawlik}, {Dickinson}, {Ferguson}, {Finlator}, {Koekemoer}, {Giavalisco},
  {Cooray}, {Dunlop}, {Faber}, {Grogin}, {Kocevski}, \&
  {Newman}}]{Finkelstein2012}
{Finkelstein}, S.~L., {Papovich}, C., {Ryan}, R.~E., {et~al.} 2012, \apj, 758,
  93

\bibitem[{{Foreman-Mackey} {et~al.}(2013){Foreman-Mackey}, {Hogg}, {Lang}, \&
  {Goodman}}]{Foreman2013}
{Foreman-Mackey}, D., {Hogg}, D.~W., {Lang}, D., \& {Goodman}, J. 2013, \pasp,
  125, 306

\bibitem[{{Giallongo} {et~al.}(2015){Giallongo}, {Grazian}, {Fiore}, {Fontana},
  {Pentericci}, {Vanzella}, {Dickinson}, {Kocevski}, {Castellano}, {Cristiani},
  {Ferguson}, {Finkelstein}, {Grogin}, {Hathi}, {Koekemoer}, {Newman}, \&
  {Salvato}}]{Giallongo2015}
{Giallongo}, E., {Grazian}, A., {Fiore}, F., {et~al.} 2015, \aap, 578, A83

\bibitem[{{Haardt} \& {Madau}(2012)}]{Haardt2012}
{Haardt}, F., \& {Madau}, P. 2012, \apj, 746, 125

\bibitem[{{Haiman} \& {Cen}(2002)}]{HC2002}
{Haiman}, Z., \& {Cen}, R. 2002, \apj, 578, 702

\bibitem[{{Heap} {et~al.}(2000){Heap}, {Williger}, {Smette}, {Hubeny}, {Sahu},
  {Jenkins}, {Tripp}, \& {Winkler}}]{Heap2000}
{Heap}, S.~R., {Williger}, G.~M., {Smette}, A., {et~al.} 2000, \apj, 534, 69

\bibitem[{{Hennawi} {et~al.}(2006){Hennawi}, {Prochaska}, {Burles}, {Strauss},
  {Richards}, {Schlegel}, {Fan}, {Schneider}, {Zakamska}, {Oguri}, {Gunn},
  {Lupton}, \& {Brinkmann}}]{Hennawi2006}
{Hennawi}, J.~F., {Prochaska}, J.~X., {Burles}, S., {et~al.} 2006, \apj, 651,
  61

\bibitem[{{Hogan} {et~al.}(1997){Hogan}, {Anderson}, \& {Rugers}}]{Hogan1997}
{Hogan}, C.~J., {Anderson}, S.~F., \& {Rugers}, M.~H. 1997, \aj, 113, 1495

\bibitem[{{Hui} {et~al.}(2001){Hui}, {Burles}, {Seljak}, {Rutledge}, {Magnier},
  \& {Tytler}}]{Hui2001}
{Hui}, L., {Burles}, S., {Seljak}, U., {et~al.} 2001, \apj, 552, 15

\bibitem[{{Hui} \& {Gnedin}(1997)}]{Hui1997}
{Hui}, L., \& {Gnedin}, N.~Y. 1997, \mnras, 292, 27

\bibitem[{{Hui} \& {Haiman}(2003)}]{Hui2003}
{Hui}, L., \& {Haiman}, Z. 2003, \apj, 596, 9

\bibitem[{{Khaire}(2017)}]{Khaire2017}
{Khaire}, V. 2017, ArXiv e-prints, arXiv:1702.03937

\bibitem[{{Khrykin} {et~al.}(2016){Khrykin}, {Hennawi}, {McQuinn}, \&
  {Worseck}}]{Khrykin2016}
{Khrykin}, I.~S., {Hennawi}, J.~F., {McQuinn}, M., \& {Worseck}, G. 2016, \apj,
  824, 133

\bibitem[{{La Plante} {et~al.}(2016){La Plante}, {Trac}, {Croft}, \&
  {Cen}}]{LaPlante2016}
{La Plante}, P., {Trac}, H., {Croft}, R., \& {Cen}, R. 2016, ArXiv e-prints,
  arXiv:1610.02047

\bibitem[{{Lidz} {et~al.}(2010){Lidz}, {Faucher-Gigu{\`e}re}, {Dall'Aglio},
  {McQuinn}, {Fechner}, {Zaldarriaga}, {Hernquist}, \& {Dutta}}]{Lidz2010}
{Lidz}, A., {Faucher-Gigu{\`e}re}, C.-A., {Dall'Aglio}, A., {et~al.} 2010,
  \apj, 718, 199

\bibitem[{{Lidz} {et~al.}(2006){Lidz}, {Heitmann}, {Hui}, {Habib}, {Rauch}, \&
  {Sargent}}]{Lidz2006}
{Lidz}, A., {Heitmann}, K., {Hui}, L., {et~al.} 2006, \apj, 638, 27

\bibitem[{{Lusso} {et~al.}(2015){Lusso}, {Worseck}, {Hennawi}, {Prochaska},
  {Vignali}, {Stern}, \& {O'Meara}}]{Lusso2015}
{Lusso}, E., {Worseck}, G., {Hennawi}, J.~F., {et~al.} 2015, ArXiv e-prints,
  arXiv:1503.02075

\bibitem[{{Madau} \& {Haardt}(2015)}]{Madau2015}
{Madau}, P., \& {Haardt}, F. 2015, \apjl, 813, L8

\bibitem[{{Madau} \& {Meiksin}(1994)}]{Madau1994}
{Madau}, P., \& {Meiksin}, A. 1994, \apjl, 433, L53

\bibitem[{{Madau} {et~al.}(2004){Madau}, {Rees}, {Volonteri}, {Haardt}, \&
  {Oh}}]{Madau2004}
{Madau}, P., {Rees}, M.~J., {Volonteri}, M., {Haardt}, F., \& {Oh}, S.~P. 2004,
  \apj, 604, 484

\bibitem[{{McDonald} {et~al.}(2001){McDonald}, {Miralda-Escud{\'e}}, {Rauch},
  {Sargent}, {Barlow}, \& {Cen}}]{McDonald2001}
{McDonald}, P., {Miralda-Escud{\'e}}, J., {Rauch}, M., {et~al.} 2001, \apj,
  562, 52

\bibitem[{{McDonald} {et~al.}(2000){McDonald}, {Miralda-Escud{\'e}}, {Rauch},
  {Sargent}, {Barlow}, {Cen}, \& {Ostriker}}]{McDonald2000}
---. 2000, \apj, 543, 1

\bibitem[{{McDonald} {et~al.}(2006){McDonald}, {Seljak}, {Burles}, {Schlegel},
  {Weinberg}, {Cen}, {Shih}, {Schaye}, {Schneider}, {Bahcall}, {Briggs},
  {Brinkmann}, {Brunner}, {Fukugita}, {Gunn}, {Ivezi{\'c}}, {Kent}, {Lupton},
  \& {Vanden Berk}}]{McDonald2006}
{McDonald}, P., {Seljak}, U., {Burles}, S., {et~al.} 2006, \apjs, 163, 80

\bibitem[{{McGreer} {et~al.}(2011){McGreer}, {Mesinger}, \&
  {Fan}}]{McGreer2011}
{McGreer}, I.~D., {Mesinger}, A., \& {Fan}, X. 2011, \mnras, 415, 3237

\bibitem[{{McQuinn}(2012)}]{McQuinn2012}
{McQuinn}, M. 2012, \mnras, 426, 1349

\bibitem[{{McQuinn} {et~al.}(2009){McQuinn}, {Lidz}, {Zaldarriaga},
  {Hernquist}, {Hopkins}, {Dutta}, \& {Faucher-Gigu{\`e}re}}]{McQuinn2009}
{McQuinn}, M., {Lidz}, A., {Zaldarriaga}, M., {et~al.} 2009, \apj, 694, 842

\bibitem[{{McQuinn} \& {Switzer}(2010)}]{McQuinn2010}
{McQuinn}, M., \& {Switzer}, E.~R. 2010, \mnras, 408, 1945

\bibitem[{{McQuinn} \& {Upton Sanderbeck}(2016)}]{McQuinn2016}
{McQuinn}, M., \& {Upton Sanderbeck}, P.~R. 2016, \mnras, 456, 47

\bibitem[{{McQuinn} \& {Worseck}(2014)}]{McQuinn2014}
{McQuinn}, M., \& {Worseck}, G. 2014, \mnras, 440, 2406

\bibitem[{{Meiksin} {et~al.}(2010){Meiksin}, {Tittley}, \&
  {Brown}}]{Meiksin2010}
{Meiksin}, A., {Tittley}, E.~R., \& {Brown}, C.~K. 2010, \mnras, 401, 77

\bibitem[{{Mellema} {et~al.}(2006){Mellema}, {Iliev}, {Alvarez}, \&
  {Shapiro}}]{Mellema2006}
{Mellema}, G., {Iliev}, I.~T., {Alvarez}, M.~A., \& {Shapiro}, P.~R. 2006, NA,
  11, 374

\bibitem[{{Miniati} {et~al.}(2004){Miniati}, {Ferrara}, {White}, \&
  {Bianchi}}]{Miniati2004}
{Miniati}, F., {Ferrara}, A., {White}, S.~D.~M., \& {Bianchi}, S. 2004, \mnras,
  348, 964

\bibitem[{{Miralda-Escud{\'e}} {et~al.}(2000){Miralda-Escud{\'e}}, {Haehnelt},
  \& {Rees}}]{Miralda2000}
{Miralda-Escud{\'e}}, J., {Haehnelt}, M., \& {Rees}, M.~J. 2000, \apj, 530, 1

\bibitem[{{Miralda-Escud{\'e}} \& {Rees}(1994)}]{Miralda1994}
{Miralda-Escud{\'e}}, J., \& {Rees}, M.~J. 1994, \mnras, 266, 343

\bibitem[{{Noh} \& {McQuinn}(2014)}]{Noh2014}
{Noh}, Y., \& {McQuinn}, M. 2014, \mnras, 444, 503

\bibitem[{{O{\~n}orbe} {et~al.}(2016){O{\~n}orbe}, {Hennawi}, \&
  {Luki{\'c}}}]{Onorbe2016}
{O{\~n}orbe}, J., {Hennawi}, J.~F., \& {Luki{\'c}}, Z. 2016, ArXiv e-prints,
  arXiv:1607.04218

\bibitem[{{Palanque-Delabrouille} {et~al.}(2013){Palanque-Delabrouille},
  {Y{\`e}che}, {Borde}, {Le Goff}, {Rossi}, {Viel}, {Aubourg}, {Bailey},
  {Bautista}, {Blomqvist}, {Bolton}, {Bolton}, {Busca}, {Carithers}, {Croft},
  {Dawson}, {Delubac}, {Font-Ribera}, {Ho}, {Kirkby}, {Lee}, {Margala},
  {Miralda-Escud{\'e}}, {Muna}, {Myers}, {Noterdaeme}, {P{\^a}ris},
  {Petitjean}, {Pieri}, {Rich}, {Rollinde}, {Ross}, {Schlegel}, {Schneider},
  {Slosar}, \& {Weinberg}}]{PD2013}
{Palanque-Delabrouille}, N., {Y{\`e}che}, C., {Borde}, A., {et~al.} 2013, \aap,
  559, A85

\bibitem[{{Planck Collaboration} {et~al.}(2016){Planck Collaboration}, {Ade},
  {Aghanim}, {Arnaud}, {Ashdown}, {Aumont}, {Baccigalupi}, {Banday},
  {Barreiro}, {Bartlett}, \& et~al.}]{PC2016}
{Planck Collaboration}, {Ade}, P.~A.~R., {Aghanim}, N., {et~al.} 2016, \aap,
  594, A13

\bibitem[{{Power} {et~al.}(2009){Power}, {Wynn}, {Combet}, \&
  {Wilkinson}}]{Power2009}
{Power}, C., {Wynn}, G.~A., {Combet}, C., \& {Wilkinson}, M.~I. 2009, \mnras,
  395, 1146

\bibitem[{{Puchwein} {et~al.}(2015){Puchwein}, {Bolton}, {Haehnelt}, {Madau},
  {Becker}, \& {Haardt}}]{Puchwein2015}
{Puchwein}, E., {Bolton}, J.~S., {Haehnelt}, M.~G., {et~al.} 2015, \mnras, 450,
  4081

\bibitem[{{Ricotti} {et~al.}(2000){Ricotti}, {Gnedin}, \&
  {Shull}}]{Ricotti2000}
{Ricotti}, M., {Gnedin}, N.~Y., \& {Shull}, J.~M. 2000, \apj, 534, 41

\bibitem[{{Robertson} {et~al.}(2010){Robertson}, {Ellis}, {Dunlop}, {McLure},
  \& {Stark}}]{Robertson2010}
{Robertson}, B.~E., {Ellis}, R.~S., {Dunlop}, J.~S., {McLure}, R.~J., \&
  {Stark}, D.~P. 2010, \nat, 468, 49

\bibitem[{{Robertson} {et~al.}(2015){Robertson}, {Ellis}, {Furlanetto}, \&
  {Dunlop}}]{Robertson2015}
{Robertson}, B.~E., {Ellis}, R.~S., {Furlanetto}, S.~R., \& {Dunlop}, J.~S.
  2015, \apjl, 802, L19

\bibitem[{{Schawinski} {et~al.}(2015){Schawinski}, {Koss}, {Berney}, \&
  {Sartori}}]{Schawinski2015}
{Schawinski}, K., {Koss}, M., {Berney}, S., \& {Sartori}, L.~F. 2015, \mnras,
  451, 2517

\bibitem[{{Schawinski} {et~al.}(2010){Schawinski}, {Evans}, {Virani}, {Urry},
  {Keel}, {Natarajan}, {Lintott}, {Manning}, {Coppi}, {Kaviraj}, {Bamford},
  {J{\'o}zsa}, {Garrett}, {van Arkel}, {Gay}, \& {Fortson}}]{Schawinski2010}
{Schawinski}, K., {Evans}, D.~A., {Virani}, S., {et~al.} 2010, \apjl, 724, L30

\bibitem[{{Schaye} {et~al.}(2000){Schaye}, {Theuns}, {Rauch}, {Efstathiou}, \&
  {Sargent}}]{Schaye2000}
{Schaye}, J., {Theuns}, T., {Rauch}, M., {Efstathiou}, G., \& {Sargent},
  W.~L.~W. 2000, \mnras, 318, 817

\bibitem[{{Shull} {et~al.}(2010){Shull}, {France}, {Danforth}, {Smith}, \&
  {Tumlinson}}]{Shull2010}
{Shull}, J.~M., {France}, K., {Danforth}, C.~W., {Smith}, B., \& {Tumlinson},
  J. 2010, \apj, 722, 1312

\bibitem[{{Shull} {et~al.}(2012){Shull}, {Stevans}, \& {Danforth}}]{Shull2012}
{Shull}, J.~M., {Stevans}, M., \& {Danforth}, C.~W. 2012, \apj, 752, 162

\bibitem[{{Springel}(2005)}]{Springel2005}
{Springel}, V. 2005, \mnras, 364, 1105

\bibitem[{{Syphers} {et~al.}(2012){Syphers}, {Anderson}, {Zheng}, {Meiksin},
  {Schneider}, \& {York}}]{Syphers2012}
{Syphers}, D., {Anderson}, S.~F., {Zheng}, W., {et~al.} 2012, \aj, 143, 100

\bibitem[{{Syphers} \& {Shull}(2014)}]{Syphers2014}
{Syphers}, D., \& {Shull}, J.~M. 2014, \apj, 784, 42

\bibitem[{{Telfer} {et~al.}(2002){Telfer}, {Zheng}, {Kriss}, \&
  {Davidsen}}]{Telfer2002}
{Telfer}, R.~C., {Zheng}, W., {Kriss}, G.~A., \& {Davidsen}, A.~F. 2002, \apj,
  565, 773

\bibitem[{{Theuns} {et~al.}(2002c){Theuns}, {Bernardi}, {Frieman}, {Hewett},
  {Schaye}, {Sheth}, \& {Subbarao}}]{Theuns2002c}
{Theuns}, T., {Bernardi}, M., {Frieman}, J., {et~al.} 2002c, \apjl, 574, L111

\bibitem[{{Theuns} {et~al.}(1998){Theuns}, {Leonard}, {Efstathiou}, {Pearce},
  \& {Thomas}}]{Theuns1998}
{Theuns}, T., {Leonard}, A., {Efstathiou}, G., {Pearce}, F.~R., \& {Thomas},
  P.~A. 1998, \mnras, 301, 478

\bibitem[{{Theuns} \& {Zaroubi}(2000)}]{Theuns2000}
{Theuns}, T., \& {Zaroubi}, S. 2000, \mnras, 317, 989

\bibitem[{{Trainor} \& {Steidel}(2013)}]{Trainor2013}
{Trainor}, R., \& {Steidel}, C.~C. 2013, \apjl, 775, L3

\bibitem[{{Upton Sanderbeck} {et~al.}(2016){Upton Sanderbeck}, {D'Aloisio}, \&
  {McQuinn}}]{Upton2016}
{Upton Sanderbeck}, P.~R., {D'Aloisio}, A., \& {McQuinn}, M.~J. 2016, \mnras,
  460, 1885

\bibitem[{{Viel} {et~al.}(2009){Viel}, {Bolton}, \& {Haehnelt}}]{Viel2009}
{Viel}, M., {Bolton}, J.~S., \& {Haehnelt}, M.~G. 2009, \mnras, 399, L39

\bibitem[{{Worseck} {et~al.}(2014){Worseck}, {Prochaska}, {Hennawi}, \&
  {McQuinn}}]{Worseck2014}
{Worseck}, G., {Prochaska}, J.~X., {Hennawi}, J.~F., \& {McQuinn}, M. 2014,
  ArXiv e-prints, arXiv:1405.7405

\bibitem[{{Worseck} {et~al.}(2011){Worseck}, {Prochaska}, {McQuinn},
  {Dall'Aglio}, {Fechner}, {Hennawi}, {Reimers}, {Richter}, \&
  {Wisotzki}}]{Worseck2011}
{Worseck}, G., {Prochaska}, J.~X., {McQuinn}, M., {et~al.} 2011, \apjl, 733,
  L24

\bibitem[{{Zaldarriaga}(2002)}]{Zaldarriaga2002}
{Zaldarriaga}, M. 2002, \apj, 564, 153

\bibitem[{{Zaldarriaga} {et~al.}(2001){Zaldarriaga}, {Hui}, \&
  {Tegmark}}]{Zaldarriaga2001}
{Zaldarriaga}, M., {Hui}, L., \& {Tegmark}, M. 2001, \apj, 557, 519

\bibitem[{{Zheng} {et~al.}(2015){Zheng}, {Syphers}, {Meiksin}, {Kriss},
  {Schneider}, {York}, \& {Anderson}}]{Zheng2015}
{Zheng}, W., {Syphers}, D., {Meiksin}, A., {et~al.} 2015, \apj, 806, 142

\end{thebibliography}

\begin{appendix}
\section{Appendix A: Cross-Correlations and the Inference Test}
\label{ap:inf_test}  

\begin{figure*}[!t]
\centering
 \includegraphics[width=0.8\linewidth]{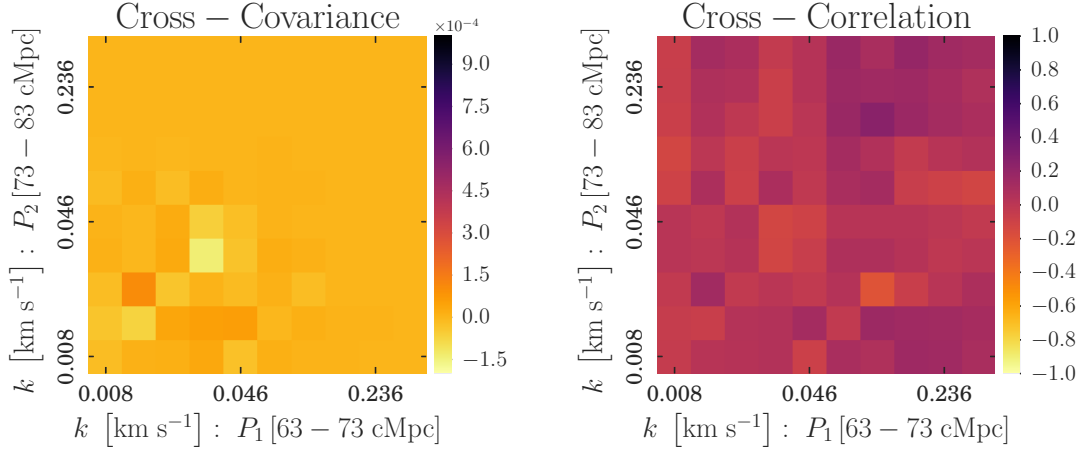}
 \caption{Cross-covariance (\emph{left panel}) and cross-correlation (\emph{right panel}) matrices for two neighboring bins: $r_1 = 63 - 73$~cMpc and $r_2 = 73-83$~cMpc. The \emph{data} sample is drawn from the model with $t_{\rm Q} = 10^8$~yr and $x_{\rm HeII,0} = 1.0$. }
\label{fig:Cross-cov}
\end{figure*}

In order to estimate the full likelihood of a given \emph{data} sample in the thermal proximity region (see Section~\ref{sec:like}), we assumed that the cross-correlation terms in the multivariate Gaussian distribution, describing the correlations between the power spectra in neighboring radial bins, are small and can safely be neglected. Thus, according to eqn.~(\ref{eqn:likelihood_full}), the full likelihood is just the product of likelihoods in each radial bin inside the thermal proximity region.

In order to validate the above assumption we calculate the cross-correlation
of the power spectrum between the neighboring radial bins and check if the cross-correlations are indeed small. For this consider two $10\ {\rm cMpc}\ r$-bins, i.e., $r_1$ and $r_2$, in which we calculate \emph{data} power spectra $P_{1} \left( k \right)$ and $P_{2} \left( k^\prime \right)$, respectively (each is the average of the same $N = 50$ skewers). The cross-correlation $\rho \left[ P_{1} \left( k \right), P_{2} \left( k^\prime \right) \right]$ between power spectra in these two radial bins is then given by 
\begin{equation}
\rho \left[ P_{1} \left( k \right), P_{2} \left( k^\prime \right) \right] = \frac{ \Sigma \left[P_{1} \left( k \right), P_{2} \left( k^\prime \right) \right]}{\sigma_{P_{1}\left( k \right)} \times \sigma_{P_{2}\left( k^\prime \right)}} 
\label{eqn:cross_cor}
\end{equation}
where $\Sigma \left[P_{1} \left( k \right), P_{2} \left( k^\prime \right) \right]$ is the cross-covariance matrix
\begin{equation}
\begin{split}
\Sigma \left[P_{1} \left( k \right), P_{2} \left( k^\prime \right) \right] = \Big \langle \left[ P_{1}\left(k|r\right) - \langle P_{1} \left(k|r\right)\rangle \right]\times \\ \times \left[ P_{2}\left(k^\prime|r\right) - \langle P_{2} \left(k^\prime|r\right) \rangle \right] \Big \rangle_{\rm N}
\end{split}
\label{eqn:cross_cov}
\end{equation}
estimated from $N = 500$ random samples of $P_{1} \left( k \right)$ and $P_{2} \left( k^\prime \right)$ with replacements, $\sigma_{P_{1}\left( k \right)}$ and $\sigma_{P_{2}\left( k^\prime \right)}$ are the diagonal elements of the respective auto-covariance matrix (in the corresponding bins $r_1$ and $r_2$) for each of the power spectra $P_{1} \left( k \right)$ and $P_{2} \left( k^\prime \right)$, respectively.

Figure~\ref{fig:Cross-cov} shows an example of cross-covariance $\Sigma \left[P_{1} \left( k \right), P_{2} \left( k^\prime \right) \right]$ and cross-correlation $\rho \left[ P_{1} \left( k \right), P_{2} \left( k^\prime \right) \right]$ matrices between \emph{data} power spectra in two radial bins: $r_1 = 63 - 73$~cMpc and $r_2 = 73-83$~cMpc. It is apparent from Figure~\ref{fig:Cross-cov} that cross-correlations between the power spectra in neighboring bins are small, i.e., $ \lesssim \pm 15-20\%$, but not exactly negligible. We note that the same behavior holds for any combination of neighboring radial bins. Therefore, in order to check the robustness of the results obtained in Section~\ref{sec:mcmc} for the covariance matrix (see eqn.~\ref{eqn:cov}) which does not include these cross-correlations we perform a simple inference test, described in what follows.

We calculate the power spectrum of the \emph{data} sample consisting,
as previously, of a realization of $N = 50$ skewers taken from the
model at $z = 3.9$ with ${\rm log}\ t_{\rm Q}^{\rm data} = 1.50$ and
$x_{\rm HeII,0}^{\rm data} = 0.50$. Following the procedure described
in Section~\ref{sec:like}, we calculate the full likelihood of this
\emph{data} sample using the covariance matrix given by
eqn.~(\ref{eqn:cov}), which ignores the cross-correlation terms. We
then run the MCMC algorithm in order to obtain the posterior
distributions of ${\rm log}\ t_{\rm Q}$ and $x_{\rm HeII,0}$. We
repeat this inference procedure for $N_{\rm samples} = 500$ different random realizations
of the $N=50$ skewers in the \emph{data} sample, and ask how often the true value of the parameters, i.e., ${\rm log}\ t_{\rm Q}^{\rm data} = 1.50$ and $x_{\rm HeII,0}^{\rm data} = 0.50$, falls inside the $68\%$ and $95\%$ contours of the MCMC posterior distributions. If our inference
is indeed robust, the fraction of realizations lying inside
these contours should correspond to the probability level of these
contours in the posterior, i.e., $\sim 68\%$ and $\sim 95\%$, respectively.
The inference test recovers the probability
to encounter true values of ${\rm log}\ t_{\rm Q}$ and $x_{\rm
  HeII,0}$ inside the $\sim 68\%$ contour is $P\left(68\% \right)
\simeq 66.4 \%$ and for $\sim 95\%$ contour is $P\left( 95\% \right)
\simeq 92.8 \%$, respectively. Thus, since these probabilities are so
close to the ideal case, the assumption that we can neglect the
cross-correlation terms holds, and demonstrates the robustness
of our likelihood, MCMC, and inference procedure.

\end{appendix}

\end{document}